\documentclass[11pt]{article}
\usepackage{geometry}                
\usepackage{graphicx}
\usepackage{rawfonts}
\usepackage{amssymb}
\usepackage{epstopdf}
\usepackage{amsmath,amsthm,amssymb}
\usepackage{mathtools}
\usepackage{fullpage}
\usepackage{color}
\usepackage[usenames,dvipsnames,svgnames]{xcolor}
\usepackage[colorlinks=true,citecolor=blue,linkcolor=red,urlcolor=blue]{hyperref}
\usepackage[font=footnotesize, labelfont=bf, margin=0.5cm]{caption}
\usepackage[labelformat=simple]{subcaption}
\usepackage{rawfonts}
\usepackage{etex}

\usepackage{ulem}

\input prepictex.tex
\input pictex.tex
\input postpictex.tex
\include{macros-pictex}

\newtheorem{theo}{Theorem}
\newtheorem{lemm}{Lemma}

\def\Pr{\noindent \textit{Proof: }}
\def\qed{$\Box$}

\def\dps{\displaystyle}

\def\sfrac#1#2{\hbox{\nor $\frac{#1}{#2}$}}
\def\Sfrac#1#2{\hbox{\large $\frac{#1}{#2}$}}

\def\Ref#1{(\ref{#1})}

\def\L{\left(} \def\R{\right)}

\usepackage{tikz}

\def\2;{\;\;}

\def\eps{\epsilon}
\def\Ref#1{(\ref{#1})}
\def\Sfrac#1#2{\hbox{\Large $\frac{#1}{#2}$}}
\def\sfrac#1#2{\hbox{\normalsize $\frac{#1}{#2}$}}
\title{Copolymeric stars adsorbed at a surface and subject to a force: a self-avoiding walk model}
\author{E.J. Janse van Rensburg\thanks{\href{mailto:rensburg@yorku.ca}{rensburg@yorku.ca}}\\
\small Department of Mathematics and Statistics\\
\small York University, Toronto, Ontario M3J 1P3, Canada
\and
S.~G.~Whittington\thanks{\href{mailto:swhittin@chem.utoronto.ca}{swhittin@chem.utoronto.ca}}\\
\small Department of Chemistry\\
\small University of Toronto, Toronto, Ontario M5S 3H6, Canada
}

\begin{document}

\maketitle

\begin{abstract}
We consider a model of star copolymers, based on self-avoiding walks, where the arms of the star can be chemically distinct.  
The copolymeric star is attached to an impenetrable surface at a vertex of unit degree
 and the different monomers constituting the star have different interaction strengths with the surface.  When the star is adsorbed at the surface it can be desorbed by applying a 
 force, either at a vertex of degree 1 or at the central vertex of the star.  We give some rigorous results about the free energy of the system and use these to establish the general form of the phase diagrams, and the orders of certain phase transitions in the system.  We also consider the special case of \textit{spiders}, \textit{ie} stars constrained to have all degree 1 vertices in the surface.
\end{abstract}


\section{Introduction}
\label{sec:Introduction}
The introduction of micro-manipulation techniques such as atomic force microscopy has allowed polymers adsorbed at a surface to be pulled off the surface in a controlled way \cite{Haupt1999,Zhang2003}.  This has led to a renewed interest in investigating how adsorbed polymers respond to a force and a variety of theoretical models have been considered.  For a review see \cite{Orlandini}.  Models based on self-avoiding walks \cite{Rensburg2015,MadrasSlade} have received particular attention.  There are some rigorous results about the self-avoiding walk model of a linear polymer, both when the force is applied at an end vertex \cite{Guttmann2014,Rensburg2013} and at the middle vertex of the walk \cite{Rensburg2017}, establishing the general form of the phase diagram.  These problems have also been investigated numerically by exact enumeration methods \cite{Guttmann2014,Mishra2005} and using Monte Carlo techniques \cite{Bradly2019b,Krawczyk2005,Krawczyk2004}.  Other polymer architectures have been considered, including star polymers \cite{Bradly2019a,BradlyOwczarek,Rensburg2018,Rensburg2019}, and other branched polymers such as combs \cite{Rensburg2019}.

The behaviour of copolymers is particularly interesting because of their application as steric stabilizers of colloidal dispersions \cite{Fleer,Napper}.  For instance a linear diblock copolymer can have one block that tends to adsorb on the surface of the colloidal particle while the other block extends into the dispersing medium.  When two colloidal particles approach one another the extended blocks lose entropy resulting in stabilization of the dispersion.  A self-avoiding walk model of linear diblock and triblock copolymers was considered in \cite{Rensburg2020}.  For a general review of the statistical mechanics of copolymers see for instance \cite{SoterosWhittington}.

\begin{figure}[t]
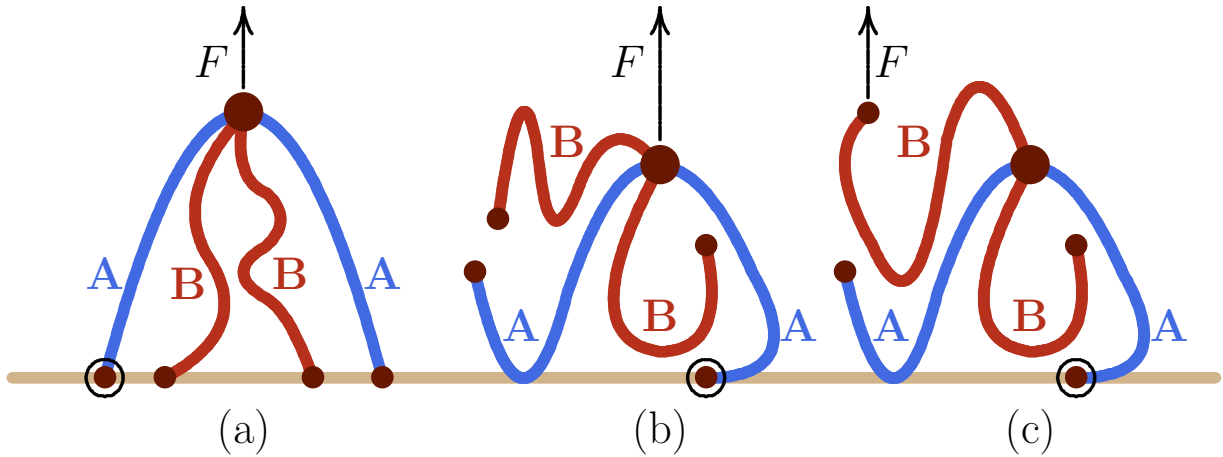

\beginpicture
\setcoordinatesystem units <1.75pt,2pt>
\setplotarea x from -30 to 240, y from -10 to 70
\setplotarea x from -20 to 240, y from -10 to 70

\setplotsymbol ({$\bullet$})
\color{Tan}
\plot -20 0 240 0 /

\setquadratic

\color{RoyalBlue}
\plot
0 0 30 50 60 0 /
\multiput {\LARGE$\mathbf{A}$} at 0 20 60 20 /

\color{BrickRed}
\plot 30 50 20 35 23 22 25 10 13 0  /
\plot 30 50 30 40 35 35 38 30 35 25 30 20 35 15 40 10 45 0 /
\multiput {\LARGE$\mathbf{B}$} at 18 18 40 20 /

\color{RoyalBlue}
\plot 80 20 90 0 100 16 110 35 120 40 130 35 140 20 144 5 130 0 /
\multiput {\LARGE$\mathbf{A}$} at 150 10 90 10 /

\color{BrickRed}
\plot 120 40 110 15 120 5 130 10 130 25 /
\plot 120 40 110 45 102 35 97 30 94 40 90 50 85 30 /
\multiput {\LARGE$\mathbf{B}$} at 120 12 100 45 /

\color{RoyalBlue}
\plot 160 20 170 0 180 16 190 35 200 40 210 35 220 20 224 5 210 0 /
\multiput {\LARGE$\mathbf{A}$} at 230 10 170 10 /

\color{BrickRed}
\plot 200 40 190 15 200 5 210 10 210 25 /
\plot 200 40 190 55 182 45 175 20 166 25 160 40 165 50 /
\multiput {\LARGE$\mathbf{B}$} at 200 12 175 45 /

\color{Sepia}
\multiput {\scalebox{2.0}{$\bullet$}} at 
0 0 60 0 13 0 45 0 80 20 130 0 130 25 85 30 160 20 210 0 210 25 165 50  /
\multiput {\scalebox{3.5}{$\bullet$}} at 30 50 120 40 200 40  /

\color{black} \normalcolor

\setplotsymbol ({$\cdot$})
\circulararc 360 degrees from 4 0 center at 0 0 
\circulararc 360 degrees from 134 0 center at 130 0 
\circulararc 360 degrees from 214 0 center at 210 0 

\arrow <10pt> [.2,.67] from 30 55 to 30 70 
\arrow <10pt> [.2,.67] from 120 45 to 120 70 
\arrow <10pt> [.2,.67] from 165 53 to 165 70 

\multiput {\LARGE$F$} at 23 60 113 60 170 60 /
\put {\LARGE(a)} at 30 -10
\put {\LARGE(b)} at 120 -10
\put {\LARGE(c)} at 200 -10
\endpicture
\caption{Schematic diagrams of adsorbing and pulled (a) copolymeric $f$-spiders, and (b) an $f$-star copolymer pulled in its central node, and (c) pulled at the endpoint of an arm. 
An endpoint of one arm is attached to the adsorbing surface at the origin, and a force $F$ is
pulling the copolymer vertically at its central node or at an endpoint of an arm.  In the case of $f$-spiders all arms are joined at the central node, and their other endpoints are constrained to be located in the adsorbing plane (and, with the exception of the endpoint of an $A$-arm at the origin, are freely moving around in the adsorbing plane).  In the case of the $f$-star copolymer, the branches may be in or disjoint with the adsorbing plane, with the exception of the endpoint of an $A$-arm which is located at the origin, as indicated.}
\label{f1}
\end{figure}

Star copolymers consist of $f$ linear polymers (called \textit{arms} or \textit{branches}) joined into a copolymer at a central node which serves as one endpoint of each linear polymer.  Star copolymers can also be used as steric stabilizers of colloidal dispersions \cite{Li} and we examine a model of copolymeric stars, based on self-avoiding walks, in this paper.  We shall focus in particular on models such as those shown in figure \ref{f1}, namely of stars pulled by forces applied at their central nodes or endpoints of arms from an adsorbing plane.  In the first model we consider stars with the endpoints of their arms constrained to be in the adsorbing plane (and one endpoint fixed at the origin in the adsorbing plane), and in the second model only one arm has an endpoint which is constrained to be fixed at the origin.

The plan of this paper is as follows.  In Section \ref{sec:review} we give a brief review of what is known rigorously for self-avoiding walks adsorbed at a surface and subject to a force.  Many of these results are used later in the paper.  A special case that will be needed later, as well as being of independent interest, is \textit{spiders} (see figure \ref{f1}(a)).  These are stars with all the vertices of degree 1 being in the surface.  These are considered in Section \ref{sec:spiders}.  We begin by considering uniform spiders pulled at their central vertex but not being attracted to the surface.  We then consider the corresponding case where the spiders are non-uniform.  We prove an upper bound on the free energy because we shall need this later.  Finally in Section \ref{sec:spiders} we look at uniform spiders, adsorbing at a surface and subject to a force that can desorb the spider. We consider the general case of stars in Section \ref{sec:stars}. For stars anchored in the surface at a vertex of degree 1, and either pulled at a vertex of degree 1 or at the central vertex, we obtain expressions for the free energy and use these, in Section \ref{sec:phases},  to deduce the principal features of the phase diagrams.  We close with a short Discussion.

\section{Notation and a brief review}
\label{sec:review}
The aim of this section is to develop some notation and to review some results about self-avoiding walks adsorbed at a surface and subject to a force.  In addition we shall mention some results about adsorbed and pulled stars and linear block copolymers.  

Consider the $d$-dimensional hypercubic lattice ${\mathbb Z}^d$, and suppose that a typical vertex has coordinates $(x_1,x_2, \ldots x_d)$.  Let $c_n$ be the number of self-avoiding walks on this lattice with $n$ edges, modulo translation.  Then we know that the limit 
\begin{equation}
\lim_{n\to\infty} \Sfrac{1}{n} \log c_n = \inf_{n>0} \Sfrac{1}{n} \log c_n = \log \mu_d
\end{equation}
exists \cite{Hammersley1957}.  $\mu_d$ is the \textit{growth constant} and $d < \mu_d < 2d-1$.  If we consider adsorption of self-avoiding walks on a $(d-1)$-dimensional hyperplane, $x_d=0$, subject to a force applied (normal to the surface) at the last vertex, we can ask for the number of self-avoiding walks starting at the origin, confined to the half-space $x_d \ge 0$ with $n$ edges, with $v+1$ vertices in the hyperplane $x_d=0$ and with the last vertex having $x_d=h$.  Write this number as $c_n(v,h)$.  We say that the walk has $v$ \textit{visits} and \textit{height} $h$.  If we weight visits with a Boltzmann factor $a= \exp[-\epsilon/kT]$ and the height with a Boltzmann factor $y=\exp[F/kT]$ the partition function is 
\begin{equation}
C_n(a,y) = \sum_v c_n(v,h)\, a^vy^h.
\label{2}
\end{equation}
Here,  $\epsilon$ is the energy associated with having a vertex in the hyperplane, $F$ is the applied force (in the vertical direction and in energy units), $k$ is Boltzmann's constant and $T$ is the absolute temperature.  In the absence of a force ($y=1$) we know that the limiting free energy
\begin{equation}
\kappa(a) = \lim_{n\to\infty} \Sfrac{1}{n} \log C_n(a,1)
\label{3}
\end{equation}
exists and there exists a critical value $a_c$ such that $\kappa(a) = \log \mu_d$ for $a \le a_c$ and $\kappa(a) > \log \mu_d$ for $a > a_c$, so that $a=a_c$ is the location of the adsorption transition \cite{HTW}.  Moreover $1 < a_c < \mu_d/\mu_{d-1}$ and $\kappa(a)$ is a convex function of $\log a$ \cite{HTW,Rensburg1998,Madras}.  Numerical estimates show that $a_c\approx 1.7756$ on the square lattice \cite{BeatonGuttmannJensen,Guttmann2014}; see also the results in reference \cite{RensburgRechnitzer}.  Similarly, if the walk does not interact with the surface ($a=1$) the free energy
\begin{equation}
\lambda(y) = \lim_{n\to\infty} \Sfrac{1}{n} \log C_n(1,y)
\label{4}
\end{equation}
exists \cite{Rensburg2009} and $\lambda(y) = \log \mu_d$ for $y \le 1$ and $\lambda(y) > \log \mu_d$ for $y > 1$ \cite{Beaton2015,IoffeVelenik,IoffeVelenik2010}.  $\lambda(y)$ is a convex function of $\log y$ \cite{Rensburg2009}.  For the general case we know that \cite{Rensburg2013}
\begin{equation}
\psi(a,y) = \lim_{n\to\infty} \Sfrac{1}{n} \log C_n(a,y) = \max[\kappa(a), \lambda(y)].
\label{5}
\end{equation}
For $a>a_c$ and $y>1$ there is a phase boundary between the adsorbed and ballistic phases given by the solution of the equation $\kappa(a) = \lambda(y)$ \cite{Rensburg2013}.

\begin{figure}[ht!]
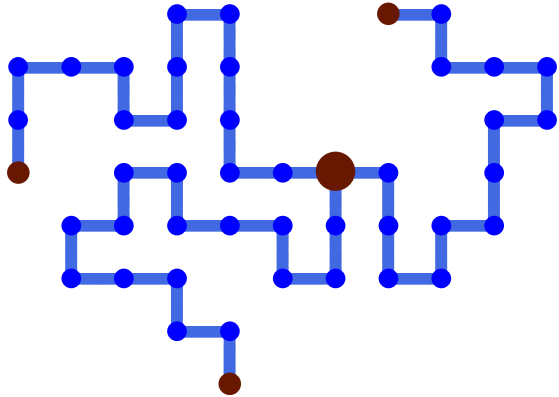

\beginpicture
\setcoordinatesystem units <2pt,2pt>
\setplotarea x from -80 to 100, y from -10 to 70
\setplotarea x from -20 to 100, y from -10 to 70

\setplotsymbol ({$\bullet$})

\color{RoyalBlue}
\plot 20 -20 20 -10 10 -10 10 0 0 0 -10 0 -10 10 0 10 0 20 10 20 10 10 20 10 30 10 30 0 40 0 40 10 40 20 /
\plot 40 20 50 20 50 10 50 0 60 0 60 10 70 10 70 20 70 30 80 30 80 40 70 40 60 40 60 50 50 50 /
\plot 40 20 30 20 20 20 20 30 20 40 20 50 10 50 10 40 10 30 0 30 0 40 -10 40 -20 40 -20 30 -20 20 /

\color{Blue}
\multiput {\scalebox{1.75}{$\bullet$}} at 
20 -20 20 -10 10 -10 10 0 0 0 -10 0 -10 10 0 10 0 20 10 20 10 10 20 10 30 10 30 0 40 0 40 10 40 20 /
\multiput {\scalebox{1.75}{$\bullet$}} at 
40 20 50 20 50 10 50 0 60 0 60 10 70 10 70 20 70 30 80 30 80 40 70 40 60 40 60 50 50 50 /
\multiput {\scalebox{1.75}{$\bullet$}} at 
40 20 30 20 20 20 20 30 20 40 20 50 10 50 10 40 10 30 0 30 0 40 -10 40 -20 40 -20 30 -20 20 /

\color{Sepia}
\multiput {\scalebox{2.0}{$\bullet$}} at 20 -20 50 50 -20 20 /
\multiput {\scalebox{3.5}{$\bullet$}} at 40 20 /

\color{black} \normalcolor

\setplotsymbol ({$\cdot$})

\endpicture
\caption{A $3$-star in the square lattice.  $f$-stars generalize in the obvious way in the hypercubic
lattice, and by allowing intersections near the central node, the numbers of arms can be increased
to an arbitrary value.}
\label{f2}
\end{figure}

 An $f$-star in the hypercubic lattice is an embedding of $f$ self-avoiding walks, all starting from the same vertex (called its \textit{central node}), and otherwise mutually avoiding (see figure \ref{f2}).  That is, an $f$-star is homeomorphic to a tree with one vertex of degree $f$, and $f$ vertices of unit degree.  If all the walks have the same length the star is \textit{uniform}.   In this paper we shall be (primarily) concerned with uniform $f$-stars in the simple cubic lattice, and so we shall often drop the adjective \textit{uniform} in what follows.  The $f$ walks are the \textit{arms} or \textit{branches} of the star.

It is obviously possible to embed $f$-stars with $2\leq f\leq 2d$ in ${\mathbb Z}^d$.  However,
uniform $f$-stars with $f>2d$ can also be treated by relaxing the self-avoiding constraint near
the central node (by allowing intersections between the arms in a small region near the central node).  
More precisely, let $r\in{\mathbb N}$ with $r \geq 1$.  If the central node of an $f$-star is located at the vertex $\vec{c}$, then the arms of the star (which are still self-avoiding walks) may mutually intersect one another in a ``ball'' $B_r(\vec{c})$ centred at $\vec{c}$ with radius $r$ (so that $B_r(\vec{c}) = \{ \vec{v}\in{\mathbb Z}^d \,|\, \| \vec{v}-\vec{c} \|_\infty \leq r \}$ using the $\ell^\infty$ norm).  Notice that each arm remains a self-avoiding walk.  If $r=1$, for example, then one can accommodate uniform $f$-stars in this way for $f\leq 9 \times 6$ in ${\mathbb Z}^3$, and $r$ grows generally proportional to $f^{1/(d-1)}$.

 In the $d$-dimensional hypercubic lattice ${\mathbb Z}^d$ for $2\leq f\leq 2d$, we write $s_n^{(f)}$ for the number of uniform $f$-stars, with $n$ edges in each arm.  In this general case it is known that \cite{WhittingtonSoteros1992}
\begin{equation}
\lim_{n\to\infty} \Sfrac{1}{nf} \log s_n^{(f)} = \log \mu_d.
\label{6}
\end{equation}
Here, $\mu_d$ is the growth constant of the self-avoiding walk.
For uniform stars with $f>2d$ the methods of reference \cite{WhittingtonSoteros1992} can be used to prove existence of the limit in equation \Ref{6} as well.  In this limit any dependence on the choice of $r$ disappears; thus, one may set \textit{a priori} a convenient value of $r$ and then work with the more general
model, accounting for possible intersections between arms near the central node.

\begin{figure}[t]
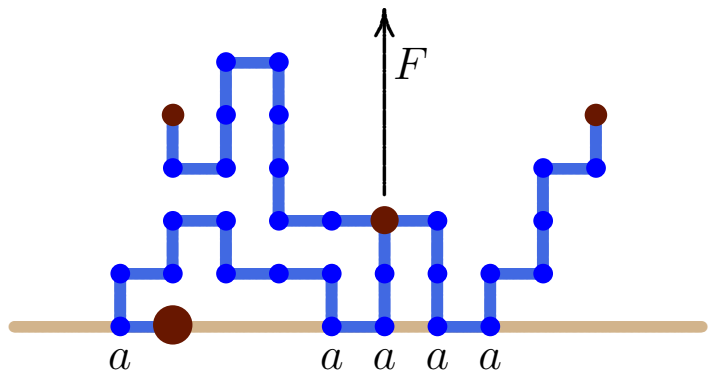

\beginpicture
\setcoordinatesystem units <2pt,2pt>
\setplotarea x from -80 to 100, y from -10 to 70
\setplotarea x from -20 to 100, y from -10 to 70

\setplotsymbol ({$\bullet$})
\color{Tan}
\plot -30 0 100 0 /

\color{RoyalBlue}
\plot 0 0 -10 0 -10 10 0 10 0 20 10 20 10 10 20 10 30 10 30 0 40 0 40 10 40 20 /
\plot 40 20 50 20 50 10 50 0 60 0 60 10 70 10 70 20 70 30 80 30 80 40 /
\plot 40 20 30 20 20 20 20 30 20 40 20 50 10 50 10 40 10 30 0 30 0 40 /

\color{Blue}
\multiput {\scalebox{1.75}{$\bullet$}} at  0 0 -10 0 -10 10 0 10 0 20 10 20 10 10 
 20 10 30 10 30 0 40 0 40 10 40 20  50 20 50 10 50 0 60 0 60 10 70 10 70 20 70 30
 80 30 80 40 30 20 20 20 20 30 20 40 20 50 10 50 10 40 10 30 0 30 0 40 /
\color{Sepia}
\multiput {\scalebox{2.0}{$\bullet$}} at 0 40 80 40  /
\multiput {\scalebox{2.5}{$\bullet$}} at 40 20 /
\multiput {\scalebox{3.5}{$\bullet$}} at 0 0  /

\color{black} \normalcolor

\setplotsymbol ({$\cdot$})
\arrow <10pt> [.2,.67] from 40 25 to 40 60 
\put {\LARGE$F$} at 45 50 
\multiput {\LARGE$a$} at -10 -6 30 -6 40 -6 50 -6 60 -6 /

\endpicture
\caption{An adsorbing $3$-star pulled with a vertical force at its central node
from the adsorbing plane.  Vertices of the star in the adsorbing plane bind
with activity $a=e^{-\eps/kT}$, where $\eps$ is a binding energy (and $k$
is Boltzmann's constant while $T$ is the absolute temperature).}
\label{f3}
\end{figure}

In this paper we continue investigating and generalizing models of adsorbing and pulled uniform cubic lattice stars introduced in references \cite{Rensburg2013,Rensburg2018}.  We considered the adsorption of $3$-stars at a surface and their desorption by an applied force.  Suppose that a $3$-star is embedded in ${\mathbb Z}^d$ with $d \geq 3$ (see figure \ref{f3}), with a vertex of degree $1$ in the hyperplane $x_d=0$, and confined to the half space $x_d \geq 0$.  Suppose that there are $v{+}1$ vertices in $x_d=0$ and suppose that the $x_d$ coordinate of another degree 1 vertex is $h$.  Write $s_n^{(3)}(v,h)$ for the number of these embeddings.  The limiting free energy (per unit length) is given by 
 \begin{equation}
 \sigma^{(3)}(a,y) = \lim_{n\to\infty} \frac{1}{3n} \log \sum_{v,h} s_n^{(3)}(v,h) a^v y^h 
 \end{equation}
 and \cite{Rensburg2018}
 \begin{equation}
 \sigma^{(3)}(a,y) = \max \left[ \kappa(a), \Sfrac{1}{3}(\lambda(y)+2\kappa(a)), \Sfrac{1}{3}(2\lambda(y) + \log \mu_d) \right].
 \end{equation} 
In addition to the phase where all arms are free (when $a < a_c$ and $y < 1$), there are three phases, one with all three arms adsorbed, one with two arms adsorbed and one arm ballistic, and a third with two arms ballistic and one arm free.  There are extra complications in the square lattice (where $d=2$) because arms can shield other arms from the surface \cite{Bradly2019a}.  The cubic lattice case (where $d=3$) has also been investigated by Monte Carlo methods \cite{BradlyOwczarek}.
 
 The adsorption and desorption of block copolymers is interesting because of their application in steric stabilization of dispersions where one type of block tends to adsorb on the surface of the colloidal particle while the other type of block is in the surrounding medium \cite{Fleer, Napper} (in which the colloidal particle is dispersed).  The properties of the adsorbed block copolymer were modelled using a self-avoiding walk model of linear diblock and triblock copolymers in reference \cite{Rensburg2020}.  The adsorbed copolymer is pulled from the surface with the force being applied at the end or at the centre of the polymer.  In reference \cite{Rensburg2020} it is shown that the phase diagrams of these models have numerous phases, including free, adsorbed, and ballistic phases while some phases have a mixed adsorbed-ballistic character.  Some of the results of this paper will be used here.

%
%
%
%
%
%

\section{Pulled Spiders}
\label{sec:spiders}

Define the \textit{half cubic lattice} 
${\mathbb Z}^3_+ = \{(x_1,x_2,x_3)\in{\mathbb Z}^3 \;\vert\; x_3 \geq 0\}$.
A \textit{spider} is a star in ${\mathbb Z}^3_+$ with the endpoints of
all its arms having height zero (and one arm has endpoint fixed at the origin).
The arms of a spider are its \textit{legs}, and the legs meet at its central
node forming the star.  An $f$-spider has $f$ legs, and a \textit{pulled} spider
has a force pulling it vertically in its central node.  A spider is uniform if
all its legs have the same length. 

We denote the number of uniform $f$-spiders with total length $nf$ and with 
the central node having height $h$ by $t_{nf}(h)$.  The partition function
of these $f$-spiders is given by
\begin{equation}
T_{nf}^{(f)}(y) = \sum_{h=0}^n t_{nf}(h)\,y^h .
\end{equation}
The limit defining the free energy of this model is known to exist \cite{Rensburg2019}
\begin{equation}
 \lim_{n\to\infty} \Sfrac{1}{nf} \log T_{nf}^{(f)} (y) = \lambda(y^{1/f})
\end{equation}
where $\lambda(y)$ is the free energy of pulled self-avoiding walks
given in equation \Ref{4}.

\begin{figure}[t]
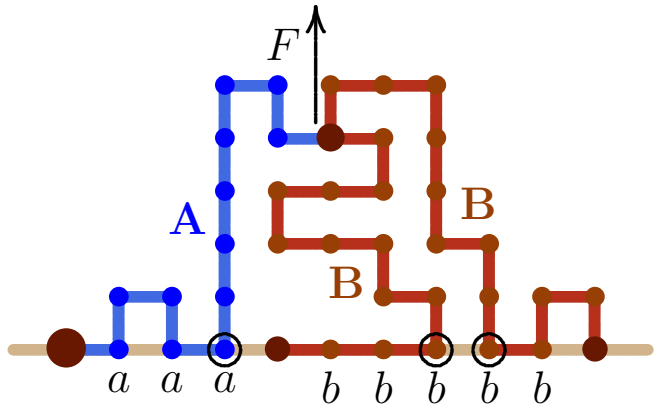

\beginpicture
\setcoordinatesystem units <2pt,2pt>
\setplotarea x from -80 to 100, y from -10 to 70
\setplotarea x from -20 to 100, y from -10 to 70

\setplotsymbol ({$\bullet$})
\color{Tan}
\plot -20 0 100 0 /

\color{RoyalBlue}
\plot -10 0 0 0 0 10 10 10 10 0 20 0 20 10 20 20 20 30 20 40 20 50 30 50 30 40 40 40  /
\color{BrickRed}
\plot 30 0 40 0 50 0 60 0 60 10 50 10 50 20 40 20 30 20 30 30 40 30 50 30 50 40 40 40  /
\plot 90 0 90 10 80 10 80 0 70 0 70 10 70 20 60 20 60 30 60 40 60 50 50 50 40 50 40 40 /

\color{Blue}
\multiput {\scalebox{1.75}{$\bullet$}} at  
0 0 0 10 10 10 10 0 20 0 20 10 20 20 20 30 20 40 20 50 30 50 30 40 40 40 /
\multiput {\LARGE$\mathbf{A}$} at 13 25 / 

\color{RawSienna}
\multiput {\scalebox{1.75}{$\bullet$}} at  
40 0 50 0 60 0 60 10 50 10 50 20 40 20 30 20 30 30 40 30 50 30 50 40 40 40  /
\multiput {\scalebox{1.75}{$\bullet$}} at  
90 0 90 10 80 10 80 0 70 0 70 10 70 20 60 20 60 30 60 40 60 50 50 50 40 50 40 40 /
\multiput {\LARGE$\mathbf{B}$} at 43 13 68 28 / 

\color{Sepia}
\multiput {\scalebox{2.25}{$\bullet$}} at 30 0 90 0 /
\multiput {\scalebox{2.5}{$\bullet$}} at 40 40 /
\multiput {\scalebox{3.5}{$\bullet$}} at -10 0  /

\color{black} \normalcolor

\setplotsymbol ({$\cdot$})
\arrow <10pt> [.2,.67] from 37.25 43 to 37.25 65 
\put {\LARGE$F$} at 31 58
\multiput {\LARGE$a$} at 0 -6 10 -6 20 -6 /
\multiput {\LARGE$b$} at 40 -7 50 -7 60 -7 70 -7 80 -7 /

\circulararc 360 degrees from 20 -3 center at 20 0
\circulararc 360 degrees from 60 -3 center at 60 0
\circulararc 360 degrees from 70 -3 center at 70 0 

\endpicture
\caption{A copolymeric $3$-spider with one $A$-leg and two $B$-legs.
The $A$ comonomers adsorb with activity $a$, and the $B$ comonomers
with activity $b$.  The spider is pulled by a vertical force applied at its
central node, while the end-points of its legs are constrained to be located
in the adsorbing plane.  One leg has an endpoint fixed at the origin (denoted
by the large bullet). Each leg is a self-avoiding walk with first vertex in the
central node, and a first visit in the adsorbing surface (marked by circles).
This first visit partitions the leg into a ballistic part with edges disjoint with the adsorbing
surface, and an adsorbing loop.}
\label{f4}
\end{figure}

In this section we consider uniform adsorbing and pulled copolymer $f$-spiders
with $g$ legs of type $A$ adsorbing with activity $a$, and $f{-}g$ legs of type $B$
adsorbing with activity $b$.  If $t_{nf}(v_a,v_b,h)$ is the number of such $f$-spiders
with $v_a$ $A$-vertices in the adsorbing plane, and $v_b$ $B$-vertices in the adsorbing
plane, with central node at height $h$, then the partition function of this model is given
by
\begin{equation}
T_{nf} (a,b,y) = \sum_{v_a,v_b} \sum_{h=0}^n t_{nf}(v_a,v_b,h)\, a^{v_a}b^{v_b} y^h .
\end{equation}
In figure \ref{f4} an example of a copolymeric $3$-spider with $f=3$ and $g=1$ is shown. 

In this section our aim is to bound the limiting free energy of this model.  Of course, the free energy of 
$f$-spiders is less than or equal to that of the same copolymeric $f$-stars (see section 
\ref{sec:stars}).

\subsection{Pulled non-uniform spiders}

Let $c_n(h) = \sum_v c_n(v,h)$ (see equation \Ref{2}) be the number of 
self-avoiding walks from the origin in ${\mathbb Z}^3_+$ of length $n$ with
last vertex of height $h$.   The limiting free energy $\lambda (y)$ of pulled walks is given in 
equation \Ref{4}, and of adsorbing walks by $\kappa(a)$ in equation \Ref{3}.
There is an $h^*$ (a function of $n$ and $y$) such that
for any $h \in \{0,1,2\ldots,n\}$,
\begin{equation}
\hbox{
$\displaystyle c_n(h)\,y^h \leq c_n(h^*)\,y^{h^*}$
and
$\displaystyle \lambda(y) = \lim_{n\to\infty} \Sfrac{1}{n} \log \L c_n(h^*)\,y^{h^*} \R .$}
\label{b11}
\end{equation}

Next, consider $f$-spiders.   Let $n > 0$ and $0 < \alpha_i < 1$ for
$i\in\{1,2,\ldots,f\}$ such that $\alpha_i \leq \alpha_{i+1}$ and 
$\sum_i \alpha_i = 1$.  We label the $f$ legs of the $f$-spider by 
$\{1,2,3,\ldots,f\}$ and these meet at the central node such that each leg
is a self-avoiding walk in ${\mathbb Z}^3_+$ from the central node
to a vertex of height zero.   If $f > 2d$ then the self-avoiding legs may 
be intersecting one another within a ball $B_r$ with center at the central node
of fixed radius $r\geq 1$ in the $\ell^\infty$-metric.
The length of the $i$-th leg is $\lfloor \alpha_i n \rfloor$ and the total length 
of the $f$-spider is $N(n)=\sum_i \lfloor \alpha_i n \rfloor$ so that 
$n{-}f \leq N(n) \leq n$.  The central node is at a height $h$ and the first leg of length 
$\lfloor \alpha_1 n \rfloor$ has its endpoint at the origin. The longest leg is denoted the 
$f$-th leg, and it has length $\lfloor \alpha_f n \rfloor$.

The number of $f$-spiders of length $N$ is denoted by
$t_N^{\alpha_1,\alpha_2,\ldots,\alpha_f}(h)$ where $h$ is the height of the
central node and where $N=N(n)=\sum_i \lfloor \alpha_i n \rfloor$.  The partition function
of pulled $f$-spiders is given by
\begin{equation}
T_N^{\alpha_1,\alpha_2,\ldots,\alpha_f}(y)
= \sum_{h=0}^{\lfloor \alpha_1 n\rfloor} t_N^{\alpha_1,\alpha_2,\ldots,\alpha_f}(h)\,y^h .
\label{b12}
\end{equation}
Notice that $0\leq h \leq \lfloor \alpha_1 n\rfloor$ since the central node
cannot have height greater than the length of the shortest arm.  For $f>2d$ the 
partition function $T_N^{\alpha_1,\alpha_2,\ldots,\alpha_f}(y)$ is also a function of $r$, 
but since $r$ is only a function of $f$, it will not change the free energy in the limit 
as $n\to\infty$.  

In the case that $\alpha_1=\alpha_2=\ldots=\alpha_f = 1/f$ we have a uniform pulled 
$f$-spider.   The methods and results in reference \cite{WhittingtonSoteros1992} show
that the free energy of uniform $f$-spiders exists and is bounded as follows:
\begin{equation}
\tau^{1/f,1/f,\ldots,1/f}(y) 
= \lim_{N\to\infty} \Sfrac{1}{N} \log T_N^{1/f,1/f,\ldots,1/f}(y)  
= \lambda(y^{1/f}) \leq \Sfrac{1}{f}(\lambda(y) + (f{-}1)\log \mu_3)
\end{equation}
since $\lambda(y)$ is a log-convex function of $y$. In this section we prove that 
this upper bound is also valid for non-uniform spiders.  The proof is by induction,
and we first consider the base case $f=2$.

Let $f=2$ so that $0 < \alpha_1 \leq \alpha_2 < 1$ where $\alpha_1+\alpha_2 = 1$ 
and $N = N(n) = \lfloor \alpha_1 n \rfloor + \lfloor \alpha_2 n \rfloor$.  The $2$-spider is then
a loop of length $N$ with endpoints in the adsorbing surface and one endpoint at the origin,
pulled by a vertical force at a node a distance $\lfloor \alpha_1 n \rfloor$ along 
the contour of the loop from the origin. The vertex where the force is applied is the 
\textit{pulled node}.  Denote by $t_N^{\alpha_1,\alpha_2}(h)$ the number of loops of length 
$N=\lfloor \alpha_1 n\rfloor + \lfloor \alpha_2 n\rfloor$, with the pulled node
a distance $\lfloor \alpha_1 n \rfloor$ along the loop from the origin at a 
height $h$ above the adsorbing plane.  The partition function of this model is given by
\begin{equation}
T_N^{\alpha_1,\alpha_2}(y)
= \sum_{h=0}^{\lfloor \alpha_1 n\rfloor} t_N^{\alpha_1,\alpha_2}(h)\, y^h .
\label{eqna9}
\end{equation}

\begin{lemm} For $0 < \alpha_1\leq \alpha_2<1$ and $\alpha_1+\alpha_2 = 1$,
$$\tau^{\alpha_1,\alpha_2}(y) 
= \limsup_{N\to\infty} \Sfrac{1}{N} \log T_N^{\alpha_1,\alpha_2}(y) 
\leq \alpha_1 \,\lambda(y) + \alpha_2 \log \mu_3
\leq \Sfrac{1}{2}(\lambda(y)+\log \mu_3),$$
where $N=\lfloor \alpha_1 n\rfloor + \lfloor \alpha_2 n \rfloor$.
\label{lemm2}
\end{lemm}

\Pr
Let $H_N$ be the value of $h$ such that $t_N^{\alpha_1,\alpha_2}(h)\, y^h $
is the largest term on the right hand side of equation \Ref{eqna9}.  Then
$H_N$ is a function of $(\alpha_1,\alpha_2,y)$ and $H_N \leq \lfloor \alpha_1 n \rfloor$.
It follows that
\begin{equation}
t_N^{\alpha_1,\alpha_2}(H_N)\,y^{H_N}
\leq T_N^{\alpha_1,\alpha_2}(y)
\leq (\lfloor \alpha_1 n \rfloor {+} 1)\, t_N^{\alpha_1,\alpha_2}(H_N)\,y^{H_N} .
\end{equation} 
Next, each loop counted by $t_N^{\alpha_1,\alpha_2}(H_N)$ can be cut 
in the pulled node into two self-avoiding walks from the adsorbing plane with
endpoints at height $H_N$.  If these two pulled walks are considered independent, 
then the last inequality becomes
\begin{equation}
T_N^{\alpha_1,\alpha_2}(y)
\leq (\lfloor \alpha_1 n \rfloor {+} 1)\, c_{\lfloor \alpha_1 n \rfloor}(H_N)\, y^{H_N} \, 
c_{\lfloor \alpha_2 n \rfloor}(H_N)  .
\end{equation}
The factors $c_{\lfloor \alpha_1 n \rfloor}(H_N) y^{H_N}$ can be bounded
above by replacing $H_N$ by $H_n^*$, the most popular value of $h$ in
$c_{\lfloor \alpha_1 n \rfloor}(h) y^{h}$.  Clearly, $H_n^* \leq \lfloor \alpha_1 n \rfloor$
and $H_n^*$ is a function of $(n,\alpha_1,y)$.  In addition, 
since $H_n^*$ is the most popular height of pulled walks of length 
$\lfloor \alpha_1 n \rfloor$, it follows that
$\lim_{N\to\infty} \sfrac{1}{N} \log (c_{\lfloor \alpha_1 n \rfloor}(H_n^*) y^{H_n^*})
= \alpha_1 \lambda(y)$.  Notice that $c_{\lfloor \alpha_2 n \rfloor}(H_n)  
\leq c_{\lfloor \alpha_2 n \rfloor}$ as well, so that the partition function can be bounded
above by
\begin{equation}
T_N^{\alpha_1,\alpha_2}(y)
\leq (\lfloor \alpha_1 n \rfloor {+} 1)\, \L c_{\lfloor \alpha_1 n \rfloor}(H_n^*)\, y^{H_n^*}\R \, 
c_{\lfloor \alpha_2 n \rfloor} .
\end{equation} 
Taking logarithms, dividing by $N$, and taking the limsup as $N\to\infty$ 
(using equation \Ref{b11}) give
\begin{equation}
\limsup_{N\to\infty} \Sfrac{1}{N} \log T_N^{\alpha_1,\alpha_2}(y)
\leq \alpha_1 \,\lambda(y) + \alpha_2 \log \mu_3.
\end{equation}
Since $\alpha_1{+}\alpha_2 = 1$, the maximum on the right hand side 
is obtained when $\alpha_1=\alpha_2=1/2$, in particular because 
$\lambda(y) \geq \log \mu_3$ and $\alpha_1 \le \alpha_2$.  In this case 
\begin{equation}
\limsup_{N\to\infty} \Sfrac{1}{N} \log T_N^{\alpha_1,\alpha_2}(y)
\leq \Sfrac{1}{2} (\lambda(y) + \log \mu_3 ).
\label{a18}
\end{equation}
This completes the proof of the lemma. \qed

Next, consider $f$-spiders with $f>2$.  Define the $\alpha_i$ as above and
assume that $\sum_i \alpha_i = 1$.  Let
$t_N^{\alpha_1,\alpha_2,\ldots,\alpha_f}(h)$ be, as before, the number
of $f$-spiders with legs of lengths $\lfloor \alpha_i n\rfloor$ and total length
$N= N(n) = \sum_{i=1}^f \lfloor \alpha_i n \rfloor$.  The partition function of
this model is defined in equation \Ref{b12}.  There is, again, a most popular
height $H_N^{(f)}$ of the pulled central node in the partition function which
is a function of $(N,n,\alpha_i,f)$ so that
\begin{equation}
t_N^{\alpha_1,\alpha_2,\ldots,\alpha_f}(H_N^{(f)})\,y^{H_N^{(f)}}
\leq T_N^{\alpha_1,\alpha_2,\ldots,\alpha_f}(y) 
\leq (\lfloor \alpha_1 n \rfloor+1)\,
t_N^{\alpha_1,\alpha_2,\ldots,\alpha_f}(H_N^{(f)})\,y^{H_N^{(f)}},
\label{b20}
\end{equation}
with the result that
\begin{equation}
\limsup_{N\to\infty} \sfrac{1}{N} \log T_N^{\alpha_1,\alpha_2,\ldots,\alpha_f}(y) 
= \limsup_{N\to\infty} \sfrac{1}{N} 
\log \left( t_N^{\alpha_1,\alpha_2,\ldots,\alpha_f}(H_N^{(f)})\,y^{H_N^{(f)}} \right) .
\label{b21}
\end{equation}
With lemma \ref{lemm2} this gives the following theorem.

\begin{theo}
Let $f\geq 2$. 
For $0 < \alpha_1\leq\alpha_2\leq\ldots\leq\alpha_f < 1$ and $\dps \sum_{i=1}^f \alpha_i = 1$,
$$ \tau^{\alpha_1,\alpha_2,\ldots,\alpha_f}(y)
= \limsup_{N\to\infty} \sfrac{1}{N} \log T_N^{\alpha_1,\alpha_2,\ldots,\alpha_f}(y) 
\leq (\alpha_1\lambda(y) + \alpha_2 \log \mu_3 + \cdots 
+\alpha_f \log \mu_3) .$$
Since $\lambda(y)\geq \log \mu_3$ and $\alpha_1 \leq 1/f$, the maximum of
the right hand side is obtained when $\alpha_1=1/f$.  In this case
$\alpha_i=1/f$ for all $i$ and this shows that
$$ \tau^{\alpha_1,\alpha_2,\ldots,\alpha_f}(y) \leq \Sfrac{1}{f}
(\lambda(y) + (f{-}1)\log \mu_3) . $$
\label{t1}
\end{theo}
\Pr
The proof for $f=2$ is given in lemma \ref{lemm2}.  Assume that the theorem
is true for $f\geq 2$, and consider the case for $f{+}1$.  By equation
\Ref{b20},
\begin{align*}
T_N^{\alpha_1,\alpha_2,\ldots,\alpha_{f+1}}(y) 
&\leq (\lfloor \alpha_1 n \rfloor+1)\,
t_N^{\alpha_1,\alpha_2,\ldots,\alpha_{f+1}}(H_N^{(f+1)})\,y^{H_N^{(f+1)}} .
\end{align*}
Cut the longest leg (with label $f{+}1$) from the $(f{+}1)$-star, put 
$M=\sum_{i=1}^f \lfloor \alpha_i n \rfloor$ (notice that 
$N=\sum_{i=1}^{f+1}\lfloor \alpha_i n \rfloor$), and use most popular heights 
to see that
\begin{align*}
T_N^{\alpha_1,\alpha_2,\ldots,\alpha_{f+1}}(y) 
&\leq (\lfloor \alpha_1 n \rfloor{+}1)\,
t_N^{\alpha_1,\alpha_2,\ldots,\alpha_{f+1}}(H_N^{(f+1)})\,y^{H_N^{(f+1)}} \cr
&\leq (\lfloor \alpha_1 n \rfloor{+}1)\,
\L t_M^{\alpha_1,\alpha_2,\ldots,\alpha_f}(H_N^{(f+1)})\,y^{H_N^{(f+1)}} \R \, 
c_{\lfloor \alpha_{f+1} n \rfloor} (H_n^{(f+1)}) \cr
&\leq (\lfloor \alpha_1 n \rfloor{+}1)\,
\L t_M^{\alpha_1,\alpha_2,\ldots,\alpha_f}(H_N^{(f)})\,y^{H_N^{(f)}} \R \, 
c_{\lfloor \alpha_{f+1} n \rfloor} . \cr
\end{align*}
Take logarithms of the above, and divide by $N$.
Use the induction hypothesis (see equation \Ref{b21}) when taking the
limsup as $N\to\infty$ to see that
\begin{align*}
\limsup_{N\to\infty} \sfrac{1}{N} \log T_N^{\alpha_1,\alpha_2,\ldots,\alpha_{f+1}}(y) 
&\leq \limsup_{N\to\infty} \Sfrac{1}{N} 
\log \left( t_M^{\alpha_1,\alpha_2,\ldots,\alpha_f}(H_n^{(f)})\,y^{H_n^{(f)}} \right)  
+ \alpha_{f+1} \log \mu_3  \cr
&\leq (\alpha_1\lambda(y)+\alpha_2 \log \mu_3+\cdots+\alpha_f \log \mu_3)
+ \alpha_{f+1} \log \mu_3 .
\end{align*}
This completes the proof. \qed

\subsection{Pulled adsorbing uniform spiders}

In this section we consider the free energy of pulled and adsorbing $f$-spiders.  
The spiders will have $g$ legs of colour $A$ and $f{-}g$ legs of colour $B$.
As before, vertices in the $A$-legs adsorb with activity $a=e^{-\epsilon_a/kT}$ where 
$\epsilon_a$ is the binding energy of vertices of colour $A$, and $T$ is the absolute temperature
while $k$ is Boltzmann's constant.  Similarly, vertices in the $B$-legs adsorb with activity
$b=e^{-\epsilon_b/kT}$.  The $f$-spider is pulled at its central node by a vertical
force $f$.  We shall bound the limiting free energy of this model using the result in 
theorem \ref {t1}.

Let $T_{nf}^{(f)}(a,b,y)$ be the partition function of uniform $f$-spiders with $f$ legs, 
each of length $n$, of which $g$ are $A$-legs, and $f{-}g$ are $B$-legs, and the 
$f$-spider is pulled by a vertical force from its central node with activity $y$.  

The structure of the $f$-spider is given in figures \ref{f1}(a) and \ref{f4} (for
$f=3$).  A portion of each leg may be adsorbed, and, starting at the 
central node, each leg is partitioned at the first vertex in the adsorbing plane 
into a ballistic part of length $\ell$, and an adsorbed part of length $n{-}\ell$.  
Assume that for the $i$-th $A$-leg, $\ell= \alpha_i n$, and for the $i$-th $B$-leg,
$\ell = \beta_i n$.  Notice that $0\leq \alpha_i \leq 1$, and $0\leq \beta_i \leq 1$.

If each leg of the uniform $f$-spider is cut into its ballistic and adsorbed
parts, then the partition function is bounded from above by a pulled non-uniform
$f$-spider and a collection of $f$ adsorbing loops.  The non-uniform
$f$-spider has partition function 
$T_N^{\alpha_1,\ldots,\alpha_g,\beta_1,\ldots,\beta_{f-g}}(y)$
where $N=N(n)=\lfloor (\alpha_1+\cdots+\alpha_g+\beta_1+\cdots+\beta_{f-g})n\rfloor
\leq nf$ is the length of the pulled non-uniform $f$-spider.  The
adsorbing parts of each leg are loops with partition functions
$L_m(a)$ for $A$-branches of lengths $m=\alpha_1 n,\ldots,\alpha_g n$,
and $L_m(b)$ for $B$-branches of lengths $m=\beta_1 n,\ldots, \beta_{f-g} n$.
Here $L_m(a)$ is given by
\begin{equation}
L_m (a) = \sum_{v=0}^m \ell_m(v)\,a^v
\end{equation}
where $\ell_m(v)$ is the number of loops of length $m$ from the origin
with $v$ visits to the adsorbing plane.  The free energy of these loops is given by
$\lim_{m\to\infty} \sfrac{1}{m} \log L_m(a) = \kappa(a)$ \cite{HTW}.

Cutting the adsorbing loops from the legs of the $f$-spider leaves behind
a non-uniform pulled $f$-spider, and gives the upper bound
\begin{equation}
\prod_{i=1}^g L_{(1-\alpha_i)n}(a)
\prod_{j=1}^{f-g} L_{(1-\beta_j)n}(b)
\times
\sum_{h\geq 0} t_N^{\alpha_1,\ldots,\alpha_g,\beta_1,\ldots,\beta_{f-g}}(h)\,y^h
\end{equation}
on the partition function of $f$-spiders with the values of $\{\alpha_i,\beta_j\}$
fixed (where $N$ is as above).  

There is a most popular value of $h$ in the summation above, denoted by 
$H_N^*$ (a function of $(N,a,b,y,n,\{\alpha_i\},\{\beta_j\})$).  If 
$\gamma=\min\{\alpha_i,\beta_j\}$, then $H_N^* \leq \gamma n$.  

In addition, for given fixed values of $(N,n,a,b,y)$ there are choices of 
the $\{\alpha_i,\beta_j\}$ maximizing the bound above; denote these by 
$\alpha_i^{(n)}$ and $\beta_j^{(n)}$, which are now functions of $n$.

This gives an upper bound on the partition function 
\begin{equation}
T_{nf}^{(f)}(a,b,y) \leq 
\prod_{i=1}^g L_{(1-\alpha_i^{(n)})n}(a)
\prod_{j=1}^{f-g} L_{(1-\beta_j^{(n)})n}(b)
\times
(\gamma n {+} 1) t_N^{\alpha_1^{(n)},\ldots,\alpha_g^{(n)},\beta_1^{(n)},\ldots,\beta_{f-g}^{(n)}}(H_N^*)\,y^{H_N^*}.
\label{c24}
\end{equation}
Simplify the upper bound by assuming that the legs may intersect one another 
(that is, while each leg is a self-avoiding walk, the legs are not avoiding one another).  
Under these conditions, by symmetry, $\alpha_i^{(n)} = \alpha_j^{(n)} = \alpha^{(n)}$ 
and similarly, and $\beta_i^{(n)}=\beta_j^{(n)} = \beta^{(n)}$.  That is, the
values of the $\{\alpha_i,\beta_j\}$ are independent of the labels of the legs.

Take logarithms, divide by $nf$, and let $n\to\infty$ to take the limsup of the
left hand side. Then on the right hand side there are limiting values (along subsequences)
of $\alpha^{(n)}$ (denoted $\alpha$), and $\beta^{(n)}$ (denoted $\beta$), such that $\alpha^{(n)}\to\alpha$ and $\beta^{(n)}\to\beta$ as $n\to\infty$.  
This shows that, after using theorem \ref{t1},
\begin{align}
\limsup_{n\to\infty} \Sfrac{1}{nf} \log T_{nf}^{(f)}(a,b,y)
&\leq \max_{\alpha,\beta} \Sfrac{1}{f} \L\left.  g(1{-}\alpha) \kappa(a) 
 + (f{-}g)(1{-}\beta)  \kappa(b)\right. \right. \nonumber \\
&\hspace{5mm} + \left. \L g\alpha + (f{-}g)\beta \R \L \lambda(y) + (f{-}1)\log\mu_3 \R \R.
\label{e25}
\end{align}
Since the expression on the right is linear in both $\alpha$ and $\beta$
(for fixed values of $(a,b,y)$), the maximum is
realised for values $\alpha,\beta\in\{0,1\}$.  We now rule out the
possible maximum at $\alpha=1$ and $\beta=0$, or at $\alpha=0$ and $\beta=1$.

\begin{lemm}
If $\alpha=0$, or if $\beta=0$, then 
$\displaystyle \limsup_{n\to\infty} \Sfrac{1}{nf} \log T_{nf}^{(f)}(a,b,y) \leq
\Sfrac{1}{f} \L g\,\kappa(a) + (f{-}g)\,\kappa(b) \R $.
\label{lemma2}
\end{lemm}

\Pr Suppose that $\alpha=0$. Then the most popular height $H^*_n$ at finite
values of $n$ is bounded by $H^*_n \leq \alpha^{(n)} n$ since the minimum
value in $\{\alpha_i^{(n)},\beta_j^{(n)}\}$ is an upper bound on 
$H^*_n/n$.  Since $\alpha^{(n)}\to 0$ as $n\to\infty$, this shows that
$\limsup_{n\to\infty} (H^*_n/n) = 0$. 

Let $\epsilon>0$.  Then there exists an $N\in{\mathbb N}$ such that
$H^*_n \leq \epsilon\,n$ for all $n\geq N$.  Next, determine an upper bound
on $T_{nf}^{(f)}(a,b,y)$ by cutting the $f$-spider in its central node
into $f$ independent adsorbing walks, each of length $n$, and ending
at height $H^*_n$.   If $C_n(a)\equiv C_n(a,1)$ 
is the partition function of adsorbing self-avoiding walks from the origin 
of length $n$, and adsorption activity $a$, and $C_n(a;h)$ is the partition function 
when the walks end at height $h$, then this gives the upper bound
\begin{equation*}
T_{nf}^{(f)}(a,b,y) \leq  
\prod_{i=1}^g C_n(a;H^*_n)
\prod_{j=1}^{f-g} C_n(b;H^*_n)
\times (\gamma n + 1)\, y^{H^*_n},
\end{equation*}
Since $C_n(a;H^*_n) \leq C_n(a)$ and $H^*_n \leq \epsilon \,n$ if $n\geq N$,
this becomes
\begin{equation*}
T_{nf}^{(f)} (a,b,y) \leq
 C_n^g (a)\; C_n^{f-g}(b)\; \times (\gamma n {+} 1)\, \max\{1,y^{\eps n}\} 
\end{equation*}
for $n\geq N$. Take logarithms, divide by $nf$, and take the limsup of the left hand 
side as $n\to\infty$.  This gives
\begin{equation*}
\limsup_{n\to\infty} \Sfrac{1}{nf} \log T_{nf}^{(f)} (a,b,y)
\leq \Sfrac{1}{f} \L g\,\kappa(a) + (f{-}g)\,\kappa(\beta) \R
+ \Sfrac{1}{f} \max\{0, \eps \log y\} . 
\end{equation*}
Take $\epsilon\to 0$ to complete the proof. \qed

By Lemma \ref{lemma2} and evaluating the critical points in equation \Ref{e25},
we obtain Theorem \ref{theorem2}.

\begin{theo}
The limiting free energy of pulled adsorbing $f$-spiders with $g$ $A$-branches and
$f{-}g$ $B$-branches is bounded from above by
\[\limsup_{n\to\infty} \Sfrac{1}{nf} \log T_{nf}^{(f)}(a,b,y)
\leq \Sfrac{1}{f} \max\{g\,\kappa(a) + (f{-}g)\,\kappa(b),\lambda(y) + (f{-}1)\log \mu_3\}
\eqno \hbox{\qed} \]
\label{theorem2}
\end{theo}


\subsection{A lower bound}

A lower bound on $T_{nf}^{(f)}(a,b,y)$ can be obtained by constructing the
$f$-spider with legs quarantined in wedges \cite{HammersleyWhittington}. 

Let $\vec{c}$ be the central node of a pulled and adsorbing cubic lattice $f$-spider 
and draw an axis vertically through $\vec{c}$.  The half cubic lattice ${\mathbb Z}_+^3$
is now partitioned around this axis by cutting it into ``cake slice'' $\alpha$-wedges
with spine the vertical axis, with floor the adsorbing plane, and vertex angle about
the spine of size $\alpha = 2\pi/f$.

An $\alpha$-bridge is a doubly unfolded walk \cite{HammersleyWelsh}  in an $\alpha$-wedge with first vertex
in the spine of the wedge, and last vertex in the floor of the wedge.  If $\alpha$ is
small, then the $\alpha$-bridge may step outside the wedge.  We compensate for this
by erecting a ball $B_r(\vec{c})$ of sufficiently large and fixed radius $r>0$ in the
$\ell^\infty$-norm, and then the bridge may step outside the wedge while inside
the ball $B_r(\vec{c})$.  Since the bridge is a self-avoiding walk, it must eventually 
exit $B_r(\vec{c})$ whereafter it is quarantined inside the $\alpha$-wedge. 

Let $b_n^{(\alpha)}(v,h)$ be the number of $\alpha$-bridges of length $n$ inside an
$\alpha$-wedge, with $v$ visits to the floor of the wedge, and height of
the first vertex $h$ above the floor of the wedge.  The partition function
of these $\alpha$-bridges is
\begin{equation}
B_n^{(\alpha)}(a,y) = \sum_{v,h} b_n^{(\alpha)}(v,h)\,a^v y^h  . 
\label{e29}
\end{equation}
The limiting free energy of adsorbing and pulled $\alpha$-bridges is given by
Theorem 1 in reference \cite{Rensburg2019}, which we state here.

\begin{theo}[Theorem 1 \cite{Rensburg2019}]
For pulled and adsorbing $\alpha$-bridges
\[ \lim_{n\to\infty} \sfrac{1}{n} \log B_n^{(\alpha)}(a,y) = \psi(a,y). \eqno \hbox{\qed} \]
\end{theo}

Define the partition function of fixed height bridges by
\begin{equation}
B_n^{(\alpha)}(a;h) = \sum_v b_n^{(\alpha)}(v,h)\,a^v .
\end{equation}
Then by equation \Ref{e29}, $B_n^{(\alpha)}(a,y) = \sum_h B_n^{(\alpha)}(a;h)\,y^h$
and there is a most popular height $H^*$ which is a function of $(n,a,y,\alpha)$, such that
\begin{equation}
\lim_{n\to\infty} \sfrac{1}{n} \log B_n^{(\alpha)}(a,y) 
= \lim_{n\to\infty} \sfrac{1}{n} \log ( B_n^{(\alpha)}(a;H^*) \,y^{H^*}) 
= \psi(a,y) . 
\label{e30}
\end{equation}

In order to find a lower bound, join together $g$ $A$-bridges, and $f{-}g$ $B$-bridges 
in $\alpha$-wedges arranged around the spine, and disjoint outside the
ball $B_r(\vec{c})$, all with first vertex of the same height $h$.  This gives the lower
bound on the partition function of pulled and adsorbing $f$-spiders:
\begin{equation}
T_{nf}^{(f)}(a,b,y)
\geq \left( B_n^{(\alpha)}(a;h)\right)^g\,
        \left( B_n^{(\alpha)}(b;h)\right)^{f-g}\, y^h
= \left( B_n^{(\alpha)}(a;h)\, y^{h/g}\right)^g\,
        \left( B_n^{(\alpha)}(b;h)\right)^{f-g}
\end{equation}
for any $h\in\{0,1,2,\ldots,n\}$.
We proceed by maximizing one of the factors above by choosing specific values of
$h$.  Let $H^*_a$ be the most popular height for the adsorption activity $a$ in 
the first factor.  This gives
\begin{equation}
T_{nf}^{(f)}(a,b,y)
\geq \left( B_n^{(\alpha)}(a;H^*_a)\, y^{H^*_a/g}\right)^g\,
        \left( B_n^{(\alpha)}(b;H^*_a)\right)^{f-g} .
\end{equation}
In the limit as $n\to\infty$ the first factor on the right hand side gives
\begin{equation}
\lim_{n\to\infty} \sfrac{1}{nf} \log \left( B_n^{(\alpha)}(a;H^*_a)\, y^{H^*_a/g}\right)^g
= \Sfrac{g}{f} \, \psi(a,y^{1/g}),
\label{e34}
\end{equation}
by definition of $H_a^*$. If $\lambda(y^{1/g}) > \kappa(a)$, then 
$\psi(a,y^{1/g}) = \lambda(y^{1/g})$, and by lemma 4 in reference 
\cite{Rensburg2020}, for almost every $y > (\lambda^{-1}(\kappa(a)))^g$,
$\lim_{n\to\infty} (H_a^*/n) = y\sfrac{d}{dy} \lambda(y^{1/g}) = \beta > 0$.

On the other hand, if $\lambda(y^{1/g}) < \kappa(a)$, then the
right hand side of equation \Ref{e34} is independent of $y$, and so one
may assume, without loss of generality, that for arbitrary small $\eps>0$
there exists an $N_\eps$ such that $H_n^*< \eps n$ for all $n>N_\eps$.  
This shows that $\beta = 0$ in this case.

We now bound $B_n^{(\alpha)}(b;H^*_a)$ from below by only considering
$\alpha$-bridges which first take $H^*_a$ steps down to the adsorbing
plane, and then continuing as an adsorbing unfolded $\alpha$-loop for 
$n-H^*_a$ steps.  This shows that $B_n^{(\alpha)}(b;H^*_a)
\geq L_{n-H_a^*}^{(\alpha)}(b)$ (where $L_n^{(\alpha)} (b)$ is the number of
adsorbing unfolded loops \cite{HTW,HammersleyWelsh} in an $\alpha$-wedge of length $n$).  This gives the
result that
\begin{equation}
\liminf_{n\to\infty} \sfrac{1}{nf} \log \left( B_n^{(\alpha)}(b;H^*_a)\right)^{f-g}
\geq
\begin{cases}
\sfrac{f-g}{f}\, (1{-}\beta)\,\kappa(b), & \text{if $\lambda(y^{1/g}) > \kappa(a)$}; \\
\sfrac{f-g}{f}\, \kappa(b), & \text{if $\lambda(y^{1/g}) < \kappa(a)$},
\end{cases}
\end{equation}
where we recall that $\beta = y\sfrac{d}{dy} \lambda(y^{1/g})$ and
$\beta=0$ if $\lambda(y^{1/g}) < \kappa(a)$ and otherwise is positive.
Putting this together with equation \Ref{e34} then gives the lower bound
\begin{equation}
\liminf_{n\to\infty} \sfrac{1}{nf} \log T_{nf}^{(f)}(a,b,y)
\geq \Sfrac{1}{f}
\left( g\,\psi(a,y^{1/g}) + (f{-}g)(1{-}\beta)\kappa(b)) \right) .
\end{equation}
By exchanging $g$ and $f{-}g$, and $a$ and $b$, the following
\begin{equation}
\liminf_{n\to\infty} \sfrac{1}{nf} \log T_{nf}^{(f)}(a,b,y)
\geq \Sfrac{1}{f}
\left( (f{-}g)\,\psi(b,y^{1/(f-g)}) + g(1{-}\beta^\prime)\kappa(a)) \right) 
\end{equation}
where $\beta^\prime = y\sfrac{d}{dy} \lambda(y^{1/(f-g)})$,
and $\beta^\prime = 0$ if $\lambda(y^{1/(f-g)}) < \kappa(b)$.

We put the above together in the following theorem.
\begin{theo}
The limiting free energy of pulled adsorbing $f$-spiders with $g$ $A$-branches and
$f{-}g$ $B$-branches is bounded from below by
\[\liminf_{n\to\infty} \Sfrac{1}{nf} \log T_{nf}^{(f)}(a,b,y)
\geq 
\begin{cases}
\Sfrac{1}{f} \left(g\,\kappa(a)+(f{-}g)\,\kappa(b)\right),
&\hspace{-3.25cm} \text{if $\lambda(y^{1/g})\leq\kappa(a)$ and $\lambda(y^{1/(f-g)})\leq\kappa(b)$} \\
\Sfrac{1}{f} \max
\begin{cases}
g\,\lambda(y^{1/g})+(f{-}g)(1{-}\beta)\kappa(b), & \\
(f{-}g)\,\lambda(y^{1/(f-g)})+g(1{-}\beta^\prime)\kappa(a) & 
\end{cases}
& otherwise.
\end{cases}
\]
Here, $\beta=0$ if $\lambda(y^{1/g}) < \kappa(a)$ and
$\beta = y\sfrac{d}{dy} \lambda(y^{1/g})$ otherwise, and 
$\beta^\prime=0$ if $\lambda(y^{1/(f-g)}) < \kappa(b)$ and
$\beta^\prime = y\sfrac{d}{dy} \lambda(y^{1/(f-g)})$ otherwise.
\qed
\label{theorem2a}
\end{theo}
A corollary of theorems \ref{theorem2} and \ref{theorem2a} is
that for $a$ and $b$ both large enough, 
$\lim_{n\to\infty} \Sfrac{1}{nf} \log T_{nf}^{(f)}(a,b,y)
= (g\,\kappa(a)+(f{-}g)\kappa(b))/f$.  In other words, there is a fully
adsorbed phase, and the limit exists in this phase.

\color{black}

\section{Pulled adsorbing copolymeric stars}
\label{sec:stars}

\begin{figure}[t]
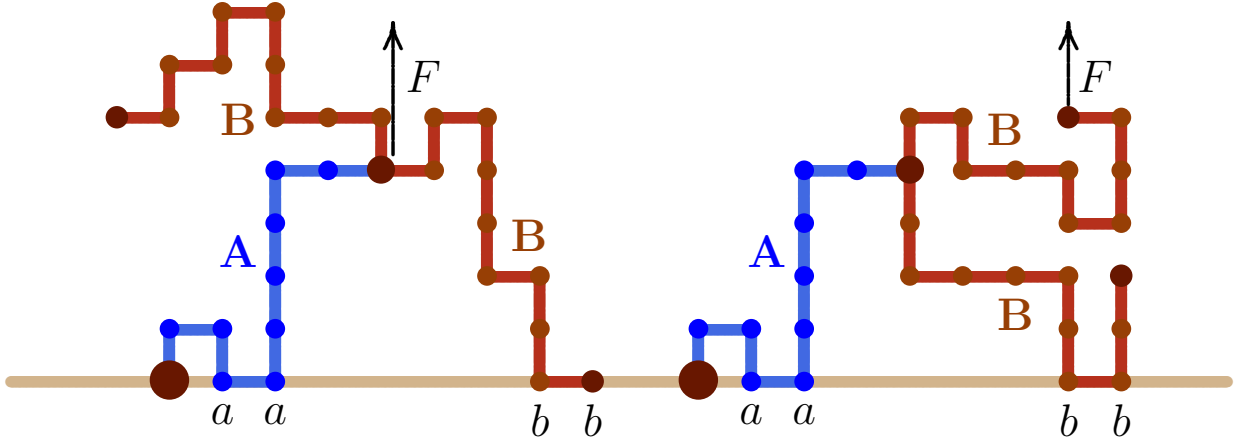

\beginpicture
\setcoordinatesystem units <2pt,2pt>
\setplotarea x from -30 to 100, y from -10 to 70
\setplotarea x from 0 to 100, y from -10 to 70

\setplotsymbol ({$\bullet$})
\color{Tan}
\plot -30 0 200 0 /

\color{RoyalBlue}
\plot 0 0 0 10 10 10 10 0 20 0 20 10 20 20 20 30 20 40 30 40 40 40  /
\color{BrickRed}
\plot 40 40 40 50 30 50 20 50 20 60 20 70 10 70 10 60 0 60 0 50 -10 50  /
\plot 80 0 70 0 70 10 70 20 60 20 60 30 60 40 60 50 50 50 50 40 40 40 /

\color{Blue}
\multiput {\scalebox{1.75}{$\bullet$}} at  
0 0 0 10 10 10 10 0 20 0 20 10 20 20 20 30 20 40 30 40 40 40 /
\multiput {\LARGE$\mathbf{A}$} at 13 25 / 

\color{RawSienna}
\multiput {\scalebox{1.75}{$\bullet$}} at  
40 40 40 50 30 50 20 50 20 60 20 70 10 70 10 60 0 60 0 50 -10 50  /
\multiput {\scalebox{1.75}{$\bullet$}} at  
80 0 70 0 70 10 70 20 60 20 60 30 60 40 60 50 50 50 50 40 40 40 /
\multiput {\LARGE$\mathbf{B}$} at 13 50 68 28 / 

\color{Sepia}
\multiput {\scalebox{2.0}{$\bullet$}} at -10 50 80 0 /
\multiput {\scalebox{2.5}{$\bullet$}} at 40 40 /
\multiput {\scalebox{3.5}{$\bullet$}} at 0 0  /

\color{black} \normalcolor

\setplotsymbol ({$\cdot$})
\arrow <10pt> [.2,.67] from 42.25 43 to 42.25 68 
\put {\LARGE$F$} at 48 58
\multiput {\LARGE$a$} at 10 -6 20 -6 /
\multiput {\LARGE$b$} at 70 -7 80 -7 /

\setcoordinatesystem units <2pt,2pt> point at -100 0

\setplotsymbol ({$\bullet$})

\color{RoyalBlue}
\plot 0 0 0 10 10 10 10 0 20 0 20 10 20 20 20 30 20 40 30 40 40 40 /
\color{BrickRed}
\plot 40 40 40 50 50 50 50 40 60 40 70 40 70 30 80 30 80 40 80 50 70 50 /
\plot 80 20 80 10 80 0 70 0 70 10 70 20 60 20 50 20 40 20 40 30 40 40 /

\color{Blue}
\multiput {\scalebox{1.75}{$\bullet$}} at  
0 0 0 10 10 10 10 0 20 0 20 10 20 20 20 30 20 40 30 40 40 40 /
\multiput {\LARGE$\mathbf{A}$} at 13 25 / 

\color{RawSienna}
\multiput {\scalebox{1.75}{$\bullet$}} at  
40 40 40 50 50 50 50 40 60 40 70 40 70 30 80 30 80 40 80 50 70 50  /
\multiput {\scalebox{1.75}{$\bullet$}} at  
80 20 80 10 80 0 70 0 70 10 70 20 60 20 50 20 40 20 40 30 40 40 /
\multiput {\LARGE$\mathbf{B}$} at 58 48 60 13 / 

\color{Sepia}
\multiput {\scalebox{2.0}{$\bullet$}} at 70 50 80 20 /
\multiput {\scalebox{2.5}{$\bullet$}} at 40 40 /
\multiput {\scalebox{3.5}{$\bullet$}} at 0 0  /

\color{black} \normalcolor

\setplotsymbol ({$\cdot$})
\arrow <10pt> [.2,.67] from 70 52.5 to 70 68 
\put {\LARGE$F$} at 75 58
\multiput {\LARGE$a$} at 10 -6 20 -6 /
\multiput {\LARGE$b$} at 70 -7 80 -7 /

\endpicture
\caption{Copolymeric $3$-stars with one $A$- and two $B$-arms.
The $A$ comonomers adsorb with activity $a$, and the $B$ comonomers
with activity $b$.  The endpoint of the $A$-arm is fixed at the origin (denoted
by the large bullet), while the star is pulled by a vertical force $F$ from its 
central node (left), or from the endpoint of a $B$-arm (right).}
\label{f5}
\end{figure}

In this section we consider $AB$-comonomer $f$-stars on the simple cubic lattice, 
confined to a half space with the confining plane playing the role of a surface at 
which adsorption can occur.  The $f$ arms (or branches) are all the same length 
so the stars are \textit{uniform}.  $g$ of the arms have vertices labelled $A$ 
and $f{-}g$ arms have vertices labelled $B$. We consider two cases: (a) where 
an $A$-vertex of degree $1$ is fixed in the surface and (b) where a $B$-vertex of 
degree $1$ is fixed in the surface.  The star is then pulled vertically from the
adsorbing surface by a force $F$, either at its central node, or at an end-vertex
of an $A$-arm, or at an end-vertex of a $B$-arm, in the regime where 
$A$-vertices and $B$-vertices interact differently with the confining plane.  
See figure \ref{f5}.  If the arm is pulled at an end-vertex, then by symmetry we only 
have to consider the case where the star is attached at either an $A$- or a $B$-vertex,
and pulled at an $A$-vertex.

\subsection{Copolymeric stars with every branch having at least one vertex in the surface}
\label{sec:starslikespiders}

In this section we consider a generalization of $f$-spiders.  These are $f$-stars with
every branch having at least one vertex in the surface, and pulled at the central vertex 
of degree $f$.  We have the following theorem:

\begin{theo}
The free energy of pulled adsorbing $f$-stars with $g$ $A$-branches and
$f{-}g$ $B$-branches, with every branch having at least one vertex in the surface, 
and pulled at the vertex of degree $f$  is bounded above by
$$ \Sfrac{1}{f} \max\{g\,\kappa(a) + (f{-}g)\,\kappa(b),\lambda(y) + (f{-}1)\log \mu_3\},$$
\label{theorem3}
\end{theo}
\Pr
The proof follows the same lines as that of Theorem \ref{theorem2} except that loops are replaced by positive walks.
\qed

The lower bound on $f$-spiders in theorem \ref{theorem2a} is also a lower
bound on the $f$-stars in theorem \ref{theorem3}.

\subsection{Copolymeric stars pulled at an end-vertex}
\label{sec:starsendpoint}

\begin{figure}[t]
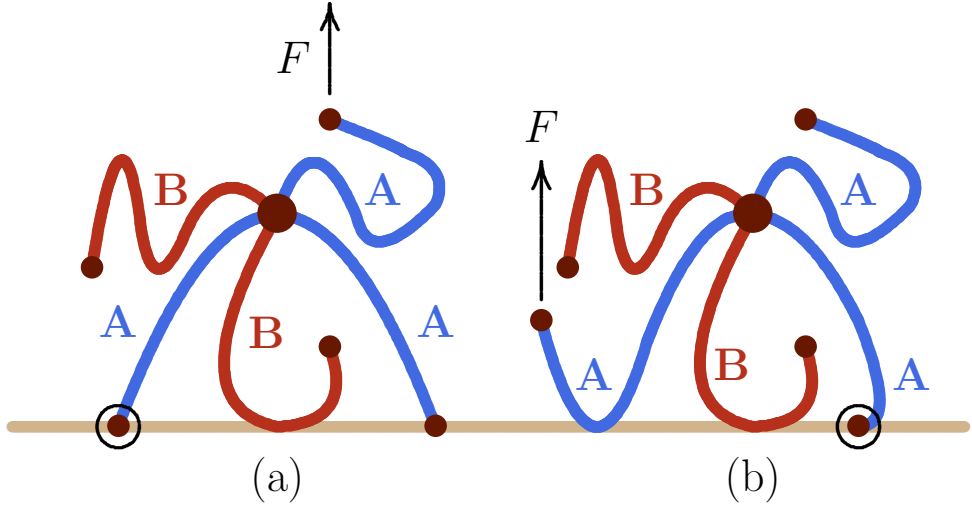

\beginpicture
\setcoordinatesystem units <2pt,2pt>
\setplotarea x from -50 to 100, y from -10 to 70
\setplotarea x from -20 to 100, y from -10 to 70

\setplotsymbol ({$\bullet$})
\color{Tan}
\plot -20 0 160 0 /

\setquadratic

\color{RoyalBlue}
\plot 0 0 30 40 60 0 /
\plot 30 40 37 50 45 40 50 35 60 42 60 48 54 52 50 54 40 58 /
\multiput {\LARGE$\mathbf{A}$} at 0 20 60 20 50 45 /
\color{BrickRed}
\plot 30 40 20 12 30 0 40 4 40 15 /
\plot 30 40 20 45 12 35 7 30 4 40 0 50 -5 30 /
\multiput {\LARGE$\mathbf{B}$} at 28 18 10 45 /

\color{RoyalBlue}
\plot 80 20 90 0 100 16 110 35 120 40 130 35 140 20 144 5 140 0 /
\plot 120 40 127 50 135 40 140 35 150 42 150 48 144 52 140 54 130 58 /
\multiput {\LARGE$\mathbf{A}$} at 150 10 90 10 140 45 /

\color{BrickRed}
\plot 120 40 110 12 120 0 130 4 130 15 /
\plot 120 40 110 45 102 35 97 30 94 40 90 50 85 30 /
\multiput {\LARGE$\mathbf{B}$} at 116 12 100 45 /

\color{Sepia}
\multiput {\scalebox{2.0}{$\bullet$}} at 
0 0 60 0 40 58  40 15 -5 30 140 0 80 20 130 58 130 15 85 30  /
\multiput {\scalebox{3.5}{$\bullet$}} at 30 40 120 40  /

\color{black} \normalcolor

\setplotsymbol ({$\cdot$})
\circulararc 360 degrees from 4 0 center at 0 0 
\circulararc 360 degrees from 144 0 center at 140 0 

\arrow <10pt> [.2,.67] from 40 63 to 40 80 
\arrow <10pt> [.2,.67] from 80 24 to 80 50 

\multiput {\LARGE$F$} at 33 70 80 57 /
\put {\LARGE(a)} at 30 -10
\put {\LARGE(b)} at 120 -10
\endpicture
\caption{The $AA$ case of a pulled adsorbing $f$-star $AB$-copolymer.
(a) The star is pulled at an endpoint of an $A$-branch with no vertices in the adsorbing surface.  The contribution to the 
free energy from this pulled branch is $\lambda(y)$.  (b) The star is pulled
at the endpoint of an $A$-branch with at least one vertex in the adsorbing plane.  In
this case the contribution to the free energy is $\psi(a,y)$.}
\label{f6}
\end{figure}

We consider the case where the force is applied, normal to the surface, 
at a vertex of degree $1$.  There are several cases to be considered:
\begin{enumerate}
\item
An $A$-vertex of degree $1$ is fixed in the surface and the force is applied 
at another $A$-vertex of degree $1$.  We call this the $AA$ case
(see figure \ref{f6}).
\item
An $A$-vertex of degree $1$ is fixed in the surface and the force is applied 
at a $B$-vertex of degree $1$.  We call this the $AB$ case.
\item
A $B$-vertex of degree $1$ is fixed in the surface and the force is applied 
at an $A$-vertex of degree $1$.  We call this the $BA$ case.
\item
A $B$-vertex of degree $1$ is fixed in the surface and the force is applied 
at an $B$-vertex of degree $1$.  We call this the $BB$ case.
\end{enumerate}

We have the following theorem.

\begin{theo}
Suppose that $y \ge 1$.  If a copolymeric $f$-star, with $g$ $A$-branches and 
$(f{-}g)$ $B$-branches,  is pulled at an $A$-vertex of degree $1$ the free energy is 
$$\sigma_e(a,b,y|AA) =  \frac{1}{f}\max
\left[ 2\lambda(y) + (f{-}2) \log \mu_3, (g{-}1) \kappa(a)+(f{-}g) \kappa(b)+\lambda(y), 
g \kappa(a)+(f{-}g) \kappa(b) \right]$$
and $ \sigma_e(a,b,y|BA) =  \sigma_e(a,b,y|AA)$.\\
If the copolymeric $f$-star is pulled at a $B$-vertex of degree $1$ the free energy is 
$$\sigma_e(a,b,y|AB) = \frac{1}{f}\max\left[
2\lambda(y) +(f{-}2) \log \mu_3, g \kappa(a)+ (f{-}g{-}1) \kappa(b) + \lambda(y),
g \kappa(a)+(f{-}g) \kappa(b)\right]$$
and $\sigma_e(a,b,y|BB) = \sigma_e(a,b,y|AB)$.
\label{thm6}
\end{theo}

\Pr
First consider the $AA$ case.
To obtain an upper bound we allow the branches of the $f$-star to intersect 
one another but not themselves, except, as before, in a ball centred
at the central node of arbitrary but fixed radius $r$.  Suppose that 
$g_A$ $A$-branches have at least one vertex in the surface and that 
$g_B$ $B$-branches have at least one vertex in the surface.
Then $1\leq g_A\leq g$ and $0\leq g_B\leq f{-}g$.

Suppose that the $A$-branch where the force is applied has no vertices in 
the surface,  as shown in case (a) of Figure \ref{f6}.  Then $1 \le g_A \le g{-}1$.
Decompose the $f$-star into a $(g_A{+}g_B)$-star with branches intersecting
the surface, an independent pulled $A$-branch not intersecting the
adsorbing surface, and $f{-}g_A{-}g_B{-}1$ independent $A$- and
$B$-branches not intersecting the adsorbing surface.

By theorem \ref{theorem3} the $(g_A{+}g_B)$-star contributes a
free energy term bounded above by 
$\max (g_A\kappa(a)+g_B\kappa(b),\lambda(y)+(g_A{+}g_B{-}1)\log \mu_3)$.
The pulled branch contributes $\lambda(y)$, and the remaining
independent branches $(f{-}g_A{-}g_B{-}1)\log \mu_3$.
Adding these together, and then maximizing over $1 \leq g_A \leq g{-}1$
and $0\leq g_B\leq f{-}g$, gives
\[ \lambda(y) + \max\left(
 (g{-}1)\kappa(a)+(f{-}g)\kappa(b),
 \lambda(y)+(f{-}2)\log\mu_3 \right). \]
This simplifies to
\[ \frac{1}{f} \max\left( \lambda(y)+(g{-}1)\kappa(a)+(f{-}g)\kappa(b),
2\lambda(y)+(f{-}2)\log\mu_3 \right) \]
after division by $f$ to obtain the intensive free energy.
 
Case (b) in figure \ref{f6} is treated in the same way.  Decomposing the
$f$-star as above now gives $2\leq g_A \leq g$ and again
$0\leq g_B \leq f{-}g$.  The pulled branch gives a contribution of
$\psi(a,y)$, and the contribution of the remaining part of the 
$(g_A{+}g_B)$-star with branches intersecting the surface is
again bounded by theorem \ref{theorem3}.  The remaining
$f{-}g_A{-}g_B$ branches not intersecting the surface give a
contribution of $(g_A{+}g_B) \log \mu_3$.  Adding together
and maximizing over $g_A$ and $g_B$ gives
\[ \frac{1}{f} \left( 
\psi(a,y) + \max\left(
(g{-}1)\kappa(a)+(f{-}g)\kappa(b),
\lambda(y)+(f{-}2)\log\mu_3 \right) \right). \]
If $\kappa(a)>\lambda(y)$ then $\psi(a,y)=\kappa(a)$ and this reduces to
$g\kappa(a)+(f{-}g)\kappa(b)$ and otherwise $\psi(a,y)=\lambda(y)$
and the bounds on case (a) are recovered.  Combining these bounds gives
\[ \frac{1}{f} \max\left( g\kappa(a)+(f{-}g)\kappa(b) , 
\lambda(y)+(g{-}1)\kappa(a)+(f{-}g)\kappa(b),
2\lambda(y)+(f{-}2)\log\mu_3 \right) \]
after division by $f$ to obtain the intensive free energy.  This completes
the upper bound for the $AA$-case.

We next consider the corresponding lower bounds and prove these by 
strategy arguments.  By monotonicity the free energy is bounded below by the free energy 
when $y=1$ and this is bounded below by the free energy of uniform spiders 
when $y=1$, \textit{ie} by $(g \kappa(a) + (f{-}g) \kappa(b))/f$.  Again by 
monotonicity the free energy is bounded below by the free energy when $a \le 1$ 
and $b \le 1$ and this is given by $(2 \lambda(y) + (f{-}2) \log \mu_3)/f$ 
(see lemma 2 in reference \cite{Rensburg2018}).

Finally, in the case that $\kappa (a) < \lambda(y)$ a lower
bound is obtained by using an $f$-spider with $g{-}1$ $A$-branches
and $f{-}g$ $B$-branches, and with branches quarantined into $f$ 
$\alpha$-wedges as in section 3.3.  One branch of the $f$-spider is replaced
by a pulled $A$-branch not intersecting the surface, and quarantined in a wedge,  with free energy $\lambda(y)$.
By theorem \ref{theorem2a} the lower bound on this conformation is
$(\lambda(y) + (g{-}1)\kappa(a) + (f{-}g)\kappa(b))/f$.  Putting this with the
lower bounds in the last paragraph gives a lower bound coinciding with the 
calculated upper bounds.

This completes the proof for the $AA$ case.

The proofs for the other cases are similar. \qed

If $y\leq 1$ then by monotonicity the free energy is bounded above by the free 
energy when $y=1$.  By theorem \ref{thm6} this is $ (g \kappa(a)+(f{-}g) \kappa(b))/f$.  
The free energy is bounded below by the value with $y=0$ and this is in turn bounded 
below by the free energy of spiders with $y=0$ (see theorem \ref{theorem2a}), 
with the same result.  Together with theorem \ref{thm6} this gives the following.

\begin{theo}
If a copolymeric $f$-star, with $g$ $A$-branches and $(f{-}g)$ $B$-branches,  
is pulled at an $A$-vertex of degree $1$ the free energy is 
$$\sigma_e(a,b,y|AA) =  \frac{1}{f}\max
\left[ 2\lambda(y) + (f{-}2) \log \mu_3, (g{-}1) \kappa(a)+(f{-}g) \kappa(b)+\lambda(y), 
g \kappa(a)+(f{-}g) \kappa(b) \right]$$
and $ \sigma_e(a,b,y|BA) =  \sigma_e(a,b,y|AA)$.\\
If the copolymeric $f$-star is pulled at a $B$-vertex of degree $1$ the free energy is 
$$\sigma_e(a,b,y|AB) = \frac{1}{f}\max\left[
2\lambda(y) +(f{-}2) \log \mu_3, g \kappa(a)+ (f{-}g{-}1) \kappa(b) + \lambda(y),
g \kappa(a)+(f{-}g) \kappa(b)\right]$$
and $\sigma_e(a,b,y|BB) = \sigma_e(a,b,y|AB)$.
\label{thm7}
\end{theo}

\subsection{Convexity of the free energy}
\label{sec:convexity}
We prove that the free is a convex function of $\log a$, $\log b$ and $\log y$.
The proof is the same for the four cases discussed above and we write $s_n(v_A,v_B,h)$ for the number of $f$-stars 
with $v_A$ $A$-visits, $v_B$ $B$-visits and height $h$ for whichever of the four cases is being 
considered. Let $$S_n(a,b,y) = \sum_{v_A,v_B,h} s_n(v_A,v_B,h) a^{v_A} b^{v_B} y^h.$$

\begin{theo}
For each of the $AA$, $AB$, $BA$ and $BB$ cases where the star is pulled at a vertex of degree 1
the free energy is a convex function of $\log a$, $\log b$ and $\log y$.
\label{theo:convex1}
\end{theo}
\Pr
Suppose that $0<p<1$.  H{\"{o}}lder's inequality shows that
$$S_n\left(a_1^p a_2^{1-p},  b_1^p b_2^{1-p}, y_1^p y_2^{1-p}\right) \le
S_n(a_1,b_1,y_1)^p S_n(a_2,b_2,y_2)^{1-p}.
$$
Taking logarithms and dividing by $n$ gives
$$\frac{1}{n} \log S_n(a_1^p a_2^{1-p},  b_1^p b_2^{1-p}, y_1^p y_2^{1-p}) \le
\frac{p}{n} \log S_n(a_1,b_1,y_1) + \frac{(1-p)}{n} \log S_n(a_2,b_2,y_2)
$$
so that $n^{-1} \log S_n(a,b,y)$ is a convex function of $\log a$, $\log b$, $\log y$.  Theorem \ref{thm7}  shows that the limit defining the free energy exists
and, since the limit of a sequence of convex functions is convex, this completes the proof.
\qed

This establishes that the free energy is convex in the 3-space and not only along coordinate directiuons.

\subsection{Copolymeric stars pulled at a mid-point}

In this section we consider the case where the force is applied at the vertex of degree $f$.  There are two cases to be considered:
\begin{enumerate}
\item[a)]
An $A$-vertex of degree 1 is fixed in the surface and the force is applied at the central node of degree $f$, this is case $A$.
\item[b)]
An $B$-vertex of degree 1 is fixed in the surface and the force is applied at the central node of degree $f$, this is case $B$.
\end{enumerate}

We consider the case of stars with $f$ branches, $g$ of which are $A$-branches and
$f{-}g$ are $B$-branches, pulled at the central vertex of degree $f$.

\begin{theo}
For an $f$-star with $g$ $A$-branches and $f{-}g$ $B$-branches, pulled at the 
central vertex with $y \ge 1$, the free energy is bounded above by
$$\frac{1}{f} \max\left( \lambda(y) + (f{-}1) \log \mu_3,  g \kappa(a) +(f{-}g)  \kappa(b) \right) .$$
\label{theorem8}
\end{theo}

\Pr
To get an upper bound we regard the branches as being allowed to intersect one 
another, but not themselves.  First consider the case where an $A$-vertex of 
degree $1$ is fixed in the surface at the origin.  We can have $1, 2, \ldots, f$ branches 
with at least one vertex in the surface.  There is always at least one $A$-branch with 
at least one vertex in the surface.
 
Suppose that there are $g_A$ $A$-branches with at least one vertex in the surface, 
then $1 \le g_A \le g$ and $g_B$ $B$-branches in the surface, $0 \le g_B \le f{-}g$.  
The $f{-}g_A{-}g_B$ branches not in the surface contribute $(f{-}g_A{-}g_B) \log \mu_3$ 
to the free energy.  By theorem \ref{theorem3} the remaining part contributes 
a term bounded above by 
$\max[\lambda(y) + (g_A{+}g_B{-}1) \log \mu_3, g_A \kappa(a) + g_B \kappa(b)]$ 
giving the upper bound
$$\frac{1}{f}
\max\left( \lambda(y)+(f{-}1) \log \mu_3, 
(g_A \kappa(a) + g_B \kappa(b) + (f{-}g_A{-}g_B) \log \mu_3 \right) .$$
This is bounded above by 
$$\frac{1}{f} 
\max\left( \lambda(y)+(f{-}1) \log \mu_3,  g  \kappa(a) + (f{-}g) \kappa(b) \right)$$
for all values of $g_A, g_B$ since $\kappa(a) \ge \log \mu_3$ and $\kappa(b) \ge \log \mu_3$.
This completes the proof for case $A$.

If the $B$-vertex of degree 1 is fixed in the surface (this is the $B$-case) the 
final upper bound is the same and proof is similar. \qed

In the event that $y\leq 1$ an upper bound on the free energy is obtained using the 
same approach as in the proof of theorem \ref{thm7}, and a lower bound is obtained
using theorem \ref{theorem2a} for spiders.  This shows that the free energy for
$y\leq 1$ is equal to $(g\kappa(a)+(f{-}g)\kappa(b))/f$.

\begin{theo}
For an $f$-star with $g$ $A$-branches and $f{-}g$ $B$-branches, pulled at 
the central vertex, the free energy is given by
$$ \sigma_m(a,b,y) = 
\frac{1}{f}\max\left( \lambda(y) + (f{-}1) \log \mu_3,  g \kappa(a) +  (f{-}g) \kappa(b) \right).
$$
\label{thm9}
\end{theo}

\Pr
We obtain lower bounds by monotonicity arguments.  

Since we are interested in $y \ge 1$ we get one lower bound by looking at $y=1$.
We can get a lower bound by considering the subset of spiders and using 
theorem \ref{theorem2a}.  This gives $(g \kappa(a) + (f-g) \kappa(b))/f$ 
as a lower bound.  

If $a \le a_c$ and $b \le a_c$ the free energy  is $(\lambda(y) + (f{-}1) \log \mu_3)/f$ and this gives 
a lower bound for all $a$ and $b$.  These bounds together with the upper bounds 
in the theorem \ref{theorem8} establish the required result. \qed

Next we state a theorem about the convexity of the free energy.

\begin{theo}
For both the $A$ and $B$ cases where the star is pulled at the central vertex
the free energy is a convex function of $\log a$, $\log b$ and $\log y$.
\label{theo:convex2}
\end{theo}
\Pr
The proof is almost identical to the proof of Theorem \ref{theo:convex1}.
\qed

Again, this establishes log-convexity in the 3-space and not only along coordinate directions.

\section{Phase diagrams}
\label{sec:phases}

The free energies $\kappa(a)$ and $\lambda(y)$ have critical points at $a=a_c>1$ 
\cite{HTW} (also see theorem 9.10 in reference \cite{Rensburg2015}) 
and at $y=y_c=1$ \cite{Beaton2015,IoffeVelenik}.  Numerical simulations of adsorbing and pulled walks
suggest that the adsorption transition at $a=a_c$ \cite{Rensburg2016}, and the ballistic transition
at $y=1$  \cite{BradlyOwczarek2022}, are continuous transitions, so that $a\sfrac{d}{da}\kappa(a)$
is a continuous function of $a$ and is singular at $a=a_c$, and $y\sfrac{d}{dy}\lambda(y)$ 
is a continuous function of $y$ which is singular at $y=1$.  It is a theorem
that $a\sfrac{d}{da}\kappa(a) = 0$ if $a\leq a_c$ and
$a\sfrac{d}{da}\kappa(a) > 0$ for almost all $a>a_c$ \cite{HTW}. (Note that $\kappa(a)$ is differentiable almost everywhere by convexity.)  
Similarly, it is a theorem
that $y\sfrac{d}{dy}\lambda(y)=0$  if $y\leq 1$ and $y\sfrac{d}{dy}\lambda(y)>0$ 
for almost all $y>1$ \cite{Beaton2015}.  These properties of $\kappa(a)$ and $\lambda(y)$
will make it possible to characterize the orders of the transitions in the phase diagrams
of pulled and adsorbing copolymer stars.  They will also, in particular, indicate the
location and presence of some critical lines in the phase diagrams.

\begin{figure}[t]
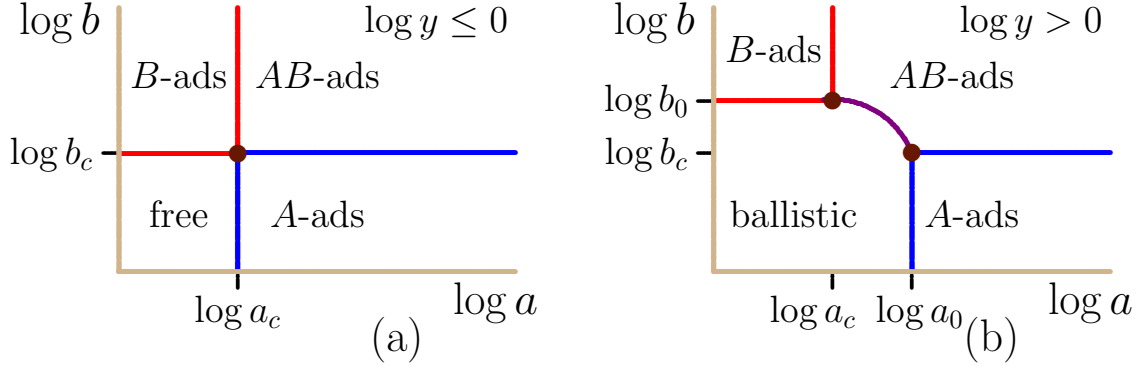

\beginpicture
\setcoordinatesystem units <1.5pt,1pt>
\setplotarea x from -30 to 100, y from -25 to 100
\color{black}

\put {\LARGE (a)} at 70 -25

\setplotsymbol ({\large$\cdot$})
\plot 0 45 -4 45 / \put {\Large$\log b_c$} at -17 45
\plot 30 0 30 -6 / \put {\Large$\log a_c$} at 30 -15
\put {\LARGE$\log a$} at 95 -11
\put {\LARGE$\log b$} at -15 95
\put {\Large{free}} at 15 22.5
\put {\Large{$B$-ads}} at 15 75
\put {\Large{$A$-ads}} at 50 22.5
\put {\Large{$AB$-ads}} at 50 75
\put {\Large{$\log y\leq 0$}} at 80 95

\setplotsymbol ({\LARGE$\cdot$})
\color{Red} 
\plot 0 45 30 45 30 100 /
\color{Blue}
\plot 30 0 30 45 100 45 /
\color{Tan}
\plot 100 0 0 0 /
\plot 0 0 0 100 /
\color{Sepia}
\put {\LARGE$\bullet$} at 30 45

\setcoordinatesystem units <1.5pt,1pt> point at -150 0
\setplotarea x from -30 to 100, y from -25 to 100
\color{black}

\put {\LARGE (b)} at 70 -25

\setplotsymbol ({\large$\cdot$})
\plot 0 45 -4 45 / \put {\Large$\log b_c$} at -17 45
\plot 30 0 30 -6 / \put {\Large$\log a_c$} at 25 -15
\plot 0 65 -4 65 / \put {\Large$\log b_0$} at -17 65
\plot 50 0 50 -6 / \put {\Large$\log a_0$} at 52 -15
\put {\LARGE$\log a$} at 95 -11
\put {\LARGE$\log b$} at -15 95
\put {\Large{ballistic}} at 20 22.5
\put {\Large{$B$-ads}} at 15 85
\put {\Large{$A$-ads}} at 65 22.5
\put {\Large{$AB$-ads}} at 60 75
\put {\Large{$\log y>0$}} at 80 95

\setplotsymbol ({\LARGE$\cdot$})
\color{Red} 
\plot 0 65 30 65 30 100 /
\color{Blue}
\plot 50 0 50 45 100 45 /
\color{Purple}
\setquadratic
\circulararc 75 degrees from 50 45 center at 30 33
\setlinear
\color{Tan}
\plot 100 0 0 0 0 100 /
\color{Sepia}
\multiput {\LARGE$\bullet$} at 30 65 50 45 /

\color{black}
\normalcolor

\endpicture
\caption{
(a) The phase diagram of an $AB$-star pulled at its central node from an adsorbing surface.  
If $y\leq 1$ then there are only free and adsorbed phases, namely a free phase
if $a\leq a_c$ and $b\leq b_c$, an $A$-adsorbed phase where the $A$-arms are adsorbed
if $a>a_c$, a similar $B$-adsorbed phase if $b>b_c$, and then an $AB$-adsorbed phase
if both $a>a_c$ and $b>b_c$.  Notice that $a_c=b_c$.
(b) If $y>1$ then the star is ballistic if it is not adsorbed. If $a$ is large and $b<b_c$, then
the star is $A$-adsorbed.  The phase boundary separating these two phases is the
solution $a=a_0$ of $g(\kappa(a)-\log\mu_3) = \lambda(y)-\log \mu_3$ for $b\leq b_c$.  
Similarly the ballistic phase and a $B$-adsorbed phase are separated by a critical value 
$b=b_0$ which is the solution of $(f{-}g)(\kappa(b)-\log\mu_3) = \lambda(y)-\log \mu_3$
for $a\leq a_c$.  An $AB$-adsorbed phase is obtained when $a>a_c$, $b>b_c$, 
and $g(\kappa(a)-\log\mu_3) + (f{-}g)(\kappa(b)-\log\mu_3) > \lambda(y)-\log \mu_3$.
Notice that $a_0>a_c$ and $b_0>b_c$ since $\lambda(y)>\log \mu_3$ if $y>1$.}
\label{f7}
\end{figure}

\subsection{Pulling at the central node}

In this model the free energy is given by theorem \ref{thm9}. Equating the values of 
the free energy in different phases gives the locations of phase boundaries. The phases 
are as follows:
\begin{enumerate}
\item
A \textit{free phase} where none of the arms are adsorbed or ballistic.  In this case, 
$a<a_c$, $b<a_c$ and $y<1$.
\item
A \textit{ballistic phase} if $\lambda(y) + (f{-}1) \log \mu_3 > g \kappa(a) + (f{-}g) \kappa(b)$.
\item
\textit{Adsorbed phases} when $\lambda(y) < g \kappa(a) + (f-g) \kappa(b) - (f-1) \log \mu_3$,
so that at least one of $a$ or $b$ must be greater than $a_c$.  Singular
points in $\kappa(a)$ and $\kappa(b)$ separate these adsorbed phases from one another.
\end{enumerate}
Phase diagrams for $y\leq 1$, and for $y>1$, are shown in figure \ref{f7}.  If
$y\leq 1$ then there are only a free phase and three adsorbed phases (namely an
$A$-adsorbed phase when $A$-arms are adsorbed, a $B$-adsorbed phase when $B$-arms
are adsorbed, and a fully $AB$-adsorbed phase when all the arms are adsorbed).  The
critical lines $a=a_c$, and $b=b_c$ are due to the singular points in $\kappa(a)$
and $\kappa(b)$ in the free energy $\sigma_m(a,b,y) = g\kappa(a) + (f{-}g)\kappa(b)$
for $y\leq 1$ in theorem \ref{thm9}. The phase transitions along the critical lines
in figure \ref{f7}(a) are of the same order as in $\kappa(a)$ or $\kappa(b)$, and
are thought to be continuous \cite{Rensburg2016}.

In the case that $y>1$ a ballistic phase appears and there are phase boundaries
separating this phase from the $A$-, $B$-, and $AB$-adsorbed phases as shown
in figure \ref{f7}.  The free energy is as follows:
\begin{equation}
\sigma_m(a,b,y) = \begin{cases}
\lambda(y) + (f{-}1)\log \mu_3, & \text{ballistic}; \\
g\,\kappa(a) + (f{-}g)\log \mu_3, & \text{$A$-adsorbed}; \\
g\,\log\mu_3 + (f{-}g)\kappa(b), & \text{$B$-adsorbed}; \\
g\,\kappa(a) + (f{-}g)\kappa(b), & \text{$AB$-adsorbed}.
\end{cases}
\end{equation}
The ballistic and $AB$-adsorbed phases are separated along a curve 
in the $ab$-plane given by the solution of 
$\lambda(y) - \log \mu_3 = g (\kappa(a) - \log \mu_3)
 + (f{-}g)(\kappa(b) - \log \mu_3)$.
 In the $ab$-plane this curve is asymptotic to $a^g\,b^{f-g} = y$.  
 
 This phase boundary is concave in the $(\log a, \log b)$-plane.
 It's convenient to write $\log a = \alpha$ and $\log b = \beta$ and the free energy as $\sigma_m(\alpha, \beta, y)$.
Throughout the ballistic phase the free energy is independent of $\alpha$ and $\beta$ so $\sigma_m(\alpha,\beta,y) = \Sigma(y)$, say.  Suppose that 
$(\alpha_1,\beta_1,y)$ and $(\alpha_2,\beta_2,y)$,  $\alpha_c \le \alpha_1, \alpha_2 \le \alpha_0$, $\beta_c \le \beta_1, \beta_2 \le \beta_0$,
are on the phase boundary of the ballistic phase, so that
$$\sigma_m(\alpha_1,\beta_1,y) = \sigma_m(\alpha_2,\beta_2,y) = \Sigma(y).$$
Since $\sigma_m$ is log-convex (see Theorem \ref{theo:convex2}) 
$$\sigma_m \left( \frac {\alpha_1+\alpha_2}{2}, \frac{\beta_1 + \beta_2}{2},y  \right) \le  \frac{ \sigma_m(\alpha_1,\beta_1,y) + \sigma_m(\alpha_2,\beta_2,y)}{2} = \Sigma(y)$$
so this point is in the ballistic phase (or on the phase boundary).  Hence, by the mid-point theorem,  the phase boundary is  concave in the $(\log a, \log b)$-plane.

The derivative of the free energy with respect to $a$  along the line $(a,b)$ with $a>0$ 
and with $0<b<b_c$ fixed is given by
\begin{equation}
a\sfrac{d}{da}\,\sigma_m(a,b,y) =
\begin{cases}
0, & \text{if $a<a_0$ and $b$ fixed in $(0,b_c)$}; \cr
g\,a\sfrac{d}{da}\, \kappa(a), & \text{if $a>a_0$ and $b$ fixed in $(0,b_c)$}.
\end{cases} 
\end{equation}
This is the mean fraction of $A$-vertices that are in the surface.
Since $a_0>a_c>1$ there is a jump discontinuity in $a\sfrac{d}{da}\,\sigma_m(a,b,y)$
as $a$ increases through $a_0$ for fixed $b\in(0,b_c)$. This shows that the phase boundary 
for $a=a_0$, $0<b<b_c$, is a first order transition.

Similarly, the phase boundary along $b=b_0$ for $0<a<a_c$ is a first order phase boundary.

The phase boundaries for $b=b_c$ and $a>a_0$, and $a=a_c$ and $b>b_0$, are
due to the singular points in $\kappa(a)$ and $\kappa(b)$ at $a=a_c$ and $b=b_c$
and so should be continuous adsorption transitions.

In the same way, the derivative of the free energy with respect to $a$ along a line $(a,b)$ with 
$a>0$ and $b$ fixed so that $b_c < b < b_0$ is given by
\begin{equation}
a\sfrac{d}{da}\,\sigma_m(a,b,y) =
\begin{cases}
0, & \text{if $(g{-}1)(\kappa(a)-\log\mu_3)+(f{-}g)(\kappa(b)-\log\mu_3) < \lambda(y)-\log\mu_3$}; \cr
g\,a\sfrac{d}{da}\, \kappa(a), & \text{otherwise}.
\end{cases} 
\label{38}
\end{equation}
Since $a\sfrac{d}{da}\, \kappa(a)>0$ for almost all $a>a_c$ and the (curved) phase boundary 
separating the ballistic and the $AB$-adsorbed phases in figure \ref{f7}(b) runs along 
a locus of points with $a>a_c$, this shows that there is a jump discontinuity
in $a\sfrac{d}{da}\,\sigma_m(a,b,y)$ as this phase boundary is crossed.
This proves that the phase boundary separating the ballistic and $AB$-adsorbed 
phases is also first order.

\begin{figure}[t]
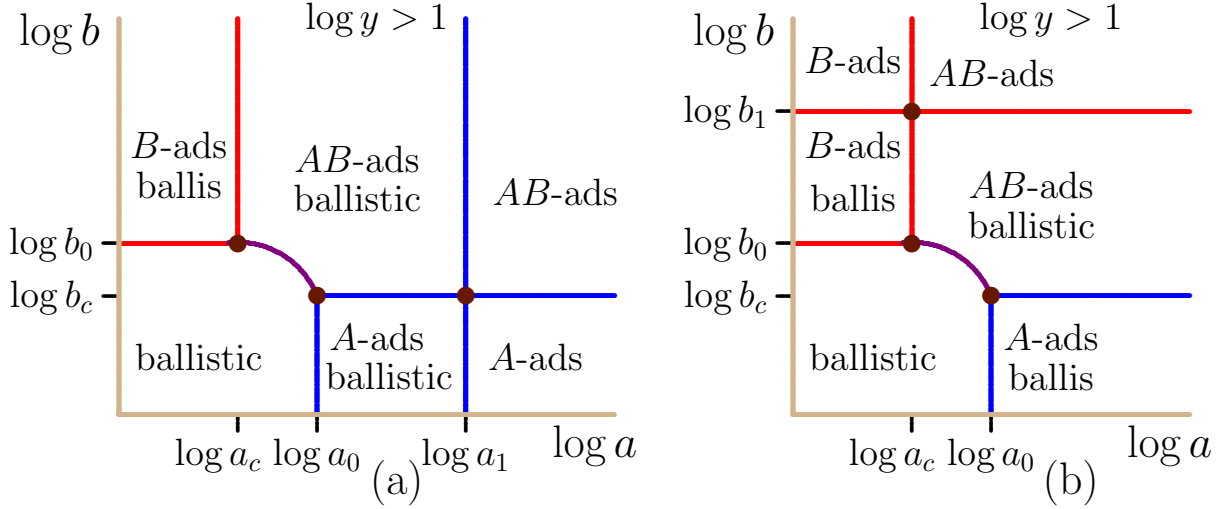

\beginpicture
\setcoordinatesystem units <1.5pt,1.pt> 
\setplotarea x from -30 to 125, y from -25 to 150
\color{black}

\put {\LARGE (a)} at 70 -25

\setplotsymbol ({\large$\cdot$})
\plot 0 45 -4 45 / \put {\Large$\log b_c$} at -17 45
\plot 30 0 30 -6 / \put {\Large$\log a_c$} at 25 -15
\plot 0 65 -4 65 / \put {\Large$\log b_0$} at -17 65
\plot 50 0 50 -6 / \put {\Large$\log a_0$} at 50 -15
\plot 87.5 0 87.5 -6 / \put {\Large$\log a_1$} at 87.5 -15
\put {\LARGE$\log a$} at 120 -11
\put {\LARGE$\log b$} at -15 145
\put {\Large{ballistic}} at 20 22.5
\put {\Large{$B$-ads}} at 15 102.5
\put {\Large{ballis}} at 15 87.5
\put {\Large{$A$-ads}} at 65 30
\put {\Large{ballistic}} at 68 15
\put {\Large{$A$-ads}} at 105 22.5
\put {\Large{$AB$-ads}} at 60 97.5
\put {\Large{ballistic}} at 60 82.5
\put {\Large{$AB$-ads}} at 110 85
\put {\Large{$\log y>1$}} at 65 150

\setplotsymbol ({\LARGE$\cdot$})
\color{Red} 
\plot 0 65 30 65 30 150 /
\color{Blue}
\plot 50 0 50 45 125 45 /
\plot 87.5 0 87.5 150 /
\color{Purple}
\setquadratic
\circulararc 75 degrees from 50 45 center at 30 33
\setlinear
\color{Tan}
\plot 125 0 0 0 0 150 /
\color{Sepia}
\multiput {\LARGE$\bullet$} at 30 65 50 45 87.5 45 /

\color{black}
\normalcolor

\setcoordinatesystem units <1.5pt,1.pt> point at -170 0
\setplotarea x from -30 to 125, y from -25 to 150
\color{black}

\put {\LARGE (b)} at 70 -25

\setplotsymbol ({\large$\cdot$})
\plot 0 45 -4 45 / \put {\Large$\log b_c$} at -17 45
\plot 30 0 30 -6 / \put {\Large$\log a_c$} at 25 -15
\plot 0 65 -4 65 / \put {\Large$\log b_0$} at -17 65
\plot 50 0 50 -6 / \put {\Large$\log a_0$} at 50 -15
\plot 0 115 -4 115 / \put {\Large$\log b_1$} at -17 115
\put {\LARGE$\log a$} at 95 -11
\put {\LARGE$\log b$} at -15 145
\put {\Large{ballistic}} at 20 22.5
\put {\Large{$B$-ads}} at 15 102.5
\put {\Large{ballis}} at 15 82.5
\put {\Large{$B$-ads}} at 15 135
\put {\Large{$A$-ads}} at 65 30
\put {\Large{ballis}} at 65 15
\put {\Large{$AB$-ads}} at 60 87.5
\put {\Large{ballistic}} at 60 72.5
\put {\Large{$AB$-ads}} at 50 130
\put {\Large{$\log y>1$}} at 65 150

\setplotsymbol ({\LARGE$\cdot$})
\color{Red} 
\plot 0 65 30 65 30 150 /
\plot 0 115 30 115 100 115 /
\color{Blue}
\plot 50 0 50 45 100 45 /
\color{Purple}
\setquadratic
\circulararc 75 degrees from 50 45 center at 30 33
\setlinear
\color{Tan}
\plot 100 0 0 0 0 150 /
\color{Sepia}
\multiput {\LARGE$\bullet$} at 30 65 50 45 30 115 /

\color{black}
\normalcolor

\endpicture
\caption{(a) The phase diagram of an $AB$-star pulled at an end-vertex of an $A$-arm
from an adsorbing surface for $y>1$.  If both $a<a_c$ and $b<b_c$, then the
star is ballistic.  Increasing $a$ while $b<b_c$ crosses a phase boundary
at $a=a_0$ into a phase which is both ballistic and $A$-adsorbed (the pulled
arm is ballistic and the remaining $A$-arms are adsorbed).  The critical point
$a_0$ is the solution of $\lambda(y) - \log \mu_3 = (g{-}1)(\kappa(a) - \log \mu_3)$.  
Further increasing $a$ takes the star through a second phase boundary at $a=a_1$
into a fully $A$-adsorbed phase.  The critical point $a_1$ is the
solution of $\kappa(a) = \lambda(y)$.
If $a<a_c$ then increasing $b$ takes the model through a phase boundary
with the $B$-arms adsorbed when $b=b_0$ (where $b_0$ is the solution
of $\lambda(y) - \log \mu_3 = (f{-}g)(\kappa(b) - \log \mu_3$).   The fully 
ballistic phase is separated from the mixed $AB$-adsorbed and ballistic phase by
a phase boundary given by the solution of 
$\lambda(y)-\log \mu_3 = (g{-}1)(\kappa(a)-\log \mu_3) + (f{-}g) (\kappa(b)-\log \mu_3)$.
The phase boundary separating the mixed $AB$-adsorbed and ballistic phase
from the fully adsorbed $AB$-adsorbed phase occurs at the solution of
$\lambda(y)=\kappa(a)$.
(b) The phase diagram if the $AB$-star is instead pulled at the endpoint of a $B$-arm.
The phases are similar, but with $a$ and $b$ interchanged.  The critical point
$b_0$ is the solution of 
$\lambda(y) - \log \mu_3 = (f{-}g{-}1)(\kappa(b) - \log \mu_3)$.  
The critical point $b_1$ is the solution of $\kappa(b)=\lambda(y)$.
The critical point $a_0$ is the solution of $\lambda(y) - \log \mu_3 = g(\kappa(a) - \log \mu_3)$.
The fully ballistic phase is separated from the mixed $AB$-adsorbed and ballistic phase
by a curved phase boundary given by the solution of
$\lambda(y)-\log \mu_3 = g(\kappa(a)-\log \mu_3) + (f{-}g{-}1) (\kappa(b)-\log \mu_3)$.}
\label{f8}
\end{figure}

\subsection{Pulling an end-vertex of an arm}

When the star is pulled at a vertex of degree $1$ at the endpoint of an arm, 
then the phase diagram depends on whether that pulled endpoint is an $A$ or $B$ vertex. 
As before, the star is rooted at the origin at an $A$-vertex of degree $1$ which is
the endpoint of an $A$-arm. 

In the case that the force is applied at an endpoint which is an $A$ vertex, then there
are the following phases:
\begin{enumerate}
\item
A \textit{free phase} where none of the branches are adsorbed or ballistic 
when $a<a_c$, $b<a_c$ and $y<1$.
\item
A \textit{ballistic phase} if $\lambda(y)-\log\mu_3 > (g{-}1)(\kappa(a)-\log\mu_3) 
+ (f{-}g)(\kappa(b) - \log \mu_3)$.
\item
\textit{Mixed ballistic and adsorbed phases} if 
$\kappa(a) < \lambda(y) < (g-1) \kappa(a) + (f-g) \kappa(b) - (f-2) \log \mu_3$.
\item
Several \textit{adsorbed phases} when $a>a_c$ and/or $b>b_c$ are sufficiently large.
\end{enumerate}
There are several phase boundaries at solutions of $\lambda(y) = \kappa(a)$ and 
$\lambda(y)-\log\mu_3 = (g{-}1)(\kappa(a)-\log\mu_3) 
 + (f{-}g)(\kappa(b) - \log \mu_3)$, as well critical lines when $\kappa(a)$ and $\kappa(b)$ 
are singular.

If $y<1$ then the phase diagram is identical to that shown in figure \ref{f7}(a).  

In the case that $y>1$, the phase diagram is shown in figure \ref{f8}(a).  Comparing
this to the phase diagram of stars pulled at the central node (see figure \ref{f7}(b))
shows the appearance of a new mixed ballistic and adsorbed phase. The free
energy is given in theorem \ref{thm7} and in the various phases is given by
\begin{equation}
\sigma_e(a,b,y|AA) = \begin{cases}
2\,\lambda(y) + (f{-}2)\log \mu_3, & \text{ballistic}; \\
\lambda(y) + (g{-}1)\,\kappa(a) + (f{-}g)\log \mu_3, & \text{$A$-ads \& ballistic}; \\
g\,\kappa(a) + (f{-}g)\log \mu_3, & \text{$A$-adsorbed}; \\
\lambda(y) + (g{-}1)\,\log\mu_3 + (f{-}g)\kappa(b), & \text{$B$-ads \& ballistic}; \\
\lambda(y) + (g{-}1)\,\kappa(a) + (f{-}g)\kappa(b), & \text{$AB$-ads \& ballistic}; \\
g\,\kappa(a) + (f{-}g)\kappa(b), & \text{$AB$-adsorbed}.
\end{cases}
\end{equation}
The critical point $a_0$ is determined by the solution of
$2\,\lambda(y) + (f{-}2)\log \mu_3 = \lambda(y) + (g{-}1)\,\kappa(a) + (f{-}g)\log \mu_3$
which simplifies to $\lambda(y)-\log\mu_3 = (g{-}1)(\kappa(a)-\log\mu_3)$.
The critical point $a_1$ is given by the solution of $\kappa(a)=\lambda(y)$.
Since $y>1$, this shows that $a_c < a_0 < a_1$.  

Fixing $b\in(0,b_c)$ and calculating the derivative of the free energy with respect to $a$ gives
\begin{equation}
a\frac{d}{da}\sigma_e(a,b,y | AA)
=\begin{cases}
0, & \text{if $0<a<a_0$}; \cr
(g{-}1)\,a\frac{d}{da}\,\kappa(a), & \text{if $a_0<a<a_1$}; \cr
g\,a\frac{d}{da}\,\kappa(a), & \text{if $a>a_1$}.
\end{cases}
\end{equation}
Since $\kappa(a)$ is strictly increasing for $a>a_c$, this shows that there are
jump discontinuities in this derivative when $a$ passes through $a_0$ and $a_1$.
That is, the phase boundaries $a=a_0$, $a=a_1$ and $0< b < b_c$ are 
first order phase boundaries.  

A similar argument shows that the phase boundary for $b=b_0$ and 
$0 < a < a_c$ is first order as well.

The phase boundary separating the $AB$-adsorbed and ballistic phase from 
the $AB$-adsorbed phase is given by the solution of
$\lambda(y) + (g{-}1)\,\kappa(a) + (f{-}g)\kappa(b)=g\,\kappa(a) + (f{-}g)\kappa(b)$
which simplifies to $\lambda(y) = \kappa(a)$.  This shows that this phase
boundary is a continuation of the phase boundary at $a=a_1$.  Taking
derivatives shows similarly that this is a first order phase boundary.

Next, consider the phase boundary if $a$ is fixed in $(a_0, a_1)$ while $b$
increases through $b_c$.  In this case the derivative with respect to $a$ of the free energy is
\begin{equation}
a\frac{d}{da}\sigma_e(a,b,y | AA)
=\begin{cases}
(g{-}1)\,a\frac{d}{da}\,\kappa(a), & \text{if $0<b<b_c$}; \cr
g\,a\frac{d}{da}\,\kappa(a), & \text{if $b>b_c$}.
\end{cases}
\end{equation}
Since $g>1$ in this model, this shows that this phase boundary is also a first
order transition.

The curved phase boundary separating the ballistic and $AB$-adsorbed phases
is given by the solution of $2\,\lambda(y) + (f{-}2)\log \mu_3 =
\lambda(y) + (g{-}1)\,\kappa(a) + (f{-}g)\log \mu_3$ which simplifies to
$\lambda(y)-\log\mu_3 = (g{-}1)(\kappa(a)-\log\mu_3) 
+ (f{-}g)(\kappa(b)-\log\mu_3)$.  This phase boundary is concave in the $(\log a, \log b)$-plane, by an argument analogous to that
given for the boundary in Figure \ref{f7}(b).  The transition across this phase boundary
is similarly a first order transition, since for $b\in(b_c,b_0)$,
\begin{equation}
a\frac{d}{da}\sigma_e(a,b,y | AA)
=\begin{cases}
0 , & \text{in the ballistic phase}; \cr
(g{-}1)\,a\frac{d}{da}\,\kappa(a), & \text{in the $AB$-ads \& ballistic phase}.
\end{cases}
\end{equation}
Since $a\frac{d}{da}\,\kappa(a)>0$ in the $AB$-adsorbed and ballistic phase
there is a jump discontinuity in the derivative along this phase boundary.
 
The remaining phase boundaries along $b=b_c$ for $a>a_1$, and for $a=a_c$ 
for $b>b_0$, are due to singular points in $\kappa(a)$ at $a=a_c$, and these
are presumably continuous transitions. 

If the force is applied at an end-vertex of a $B$-arm then the phase diagram is similar 
but the phase boundaries are in different locations (and this is shown in
figure \ref{f8}(b)).  The phases are similar, and are given by
\begin{enumerate}
\item
A \textit{free phase} where none of the branches are adsorbed or ballistic when 
$a<a_c$, $b<a_c$ and $y<1$.
\item
A \textit{ballistic phase} if $\lambda(y)-\log\mu_3 > g(\kappa(a)-\log\mu_3)
 + (f{-}g{-}1) (\kappa(b) - \log \mu_3)$.
\item
\textit{Mixed ballistic and adsorbed phase} if 
$\kappa(b) < \lambda(y) < g \kappa(a) + (f{-}g{-}1) \kappa(b) - (f{-}2) \log \mu_3$.
\item
Several \textit{adsorbed phases} when $a>a_c$ and/or $b>b_c$ are sufficiently large.
\end{enumerate}

\begin{figure}[t]
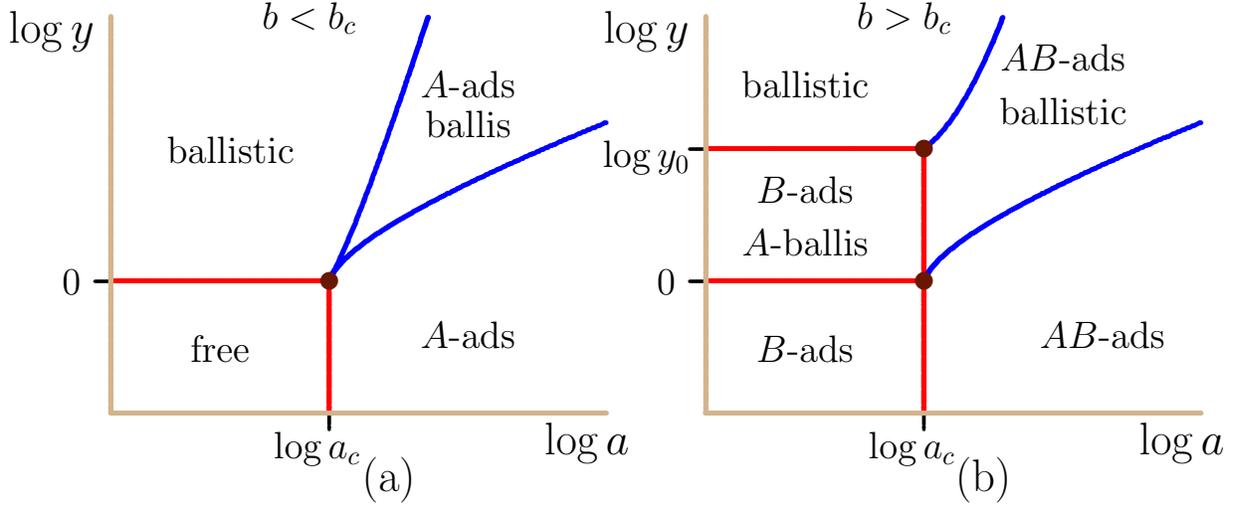

\beginpicture
\setcoordinatesystem units <1.5pt,1.pt> 
\setplotarea x from -30 to 125, y from -25 to 150
\color{black}

\put {\LARGE (a)} at 70 -25

\setplotsymbol ({\large$\cdot$})
\plot 0 50 -4 50 / \put {\Large$0$} at -10 50
\plot 55 0 55 -6 / \put {\Large$\log a_c$} at 52 -13
\put {\LARGE$\log a$} at 120 -11
\put {\LARGE$\log y$} at -15 145
\put {\Large{free}} at 27.5 25
\put {\Large{ballistic}} at 30 100
\put {\Large{$A$-ads}} at 90 125
\put {\Large{ballis}} at 90 110
\put {\Large{$A$-ads}} at 90 30
\put {\Large{$b<b_c$}} at 50 150

\setplotsymbol ({\LARGE$\cdot$})
\color{Red} 
\plot 0 50 55 50 55 0 /
\setquadratic
\color{Blue}
\plot 55 50 65 85 80 150 /
\plot 55 50 75 75 125 110 /
\setlinear
\color{Tan}
\plot 125 0 0 0 0 150 /
\color{Sepia}
\multiput {\LARGE$\bullet$} at 55 50 /

\color{black}
\normalcolor

\setcoordinatesystem units <1.5pt,1.pt> point at -150 0
\setplotarea x from -30 to 125, y from -25 to 150
\color{black}

\put {\LARGE (b)} at 70 -25

\setplotsymbol ({\large$\cdot$})
\plot 0 50 -4 50 / \put {\Large$0$} at -10 50
\plot 0 100 -4 100 / \put {\Large$\log y_0$} at -15 97
\plot 55 0 55 -6 / \put {\Large$\log a_c$} at 52 -13
\put {\LARGE$\log a$} at 120 -11
\put {\LARGE$\log y$} at -15 145
\put {\Large{ballistic}} at 25 125
\put {\Large{$B$-ads}} at 25 85
\put {\Large{$A$-ballis}} at 25 65
\put {\Large{$B$-ads}} at 25 25
\put {\Large{$AB$-ads}} at 90 135
\put {\Large{ballistic}} at 90 115
\put {\Large{$AB$-ads}} at 100 30
\put {\Large{$b>b_c$}} at 50 150

\setplotsymbol ({\LARGE$\cdot$})
\color{Red} 
\plot 0 50 55 50 55 0 /
\plot 0 100 55 100 55 50 /
\setquadratic
\color{Blue}
\plot 55 100 65 118 75 150 /
\plot 55 50 75 75 125 110 /
\setlinear
\color{Tan}
\plot 125 0 0 0 0 150 /
\color{Sepia}
\multiput {\LARGE$\bullet$} at 55 50 55 100 /

\color{black}
\normalcolor

\endpicture
\caption{Phase diagrams of the $AA$-case in the log-log $ay$-plane for fixed values
of $b$.  (a) The phase diagram for $b<b_c$.  A free phase
for small $a<a_c$ and $y<1$ gives way to a ballistic phase as $y$ is increased, and to 
an $A$-adsorbed phase as $a$ is increased through phase boundaries which
are presumably continuous.  These phases are separated from a mixed ballistic
and $A$-adsorbed phase by first order phase boundaries for large values
of $a$ and $y$.  (b) The phase diagram for $b>b_c$ fixed.  A $B$-adsorbed
phase for small $a<a_c$ and $y<1$ goes through (presumably) continuous transitions
into a mixed ballistic and $B$-adsorbed phase to a fully ballistic phase as $y$ is 
increased through $1$ and $y_0$.  Increasing $a$ instead takes the model 
through a (presumably) continuous phase transition to an $AB$-adsorbed phase 
at $a=a_c$.  The ballistic phase is separated from a mixed ballistic and
$AB$-adsorbed phase through a first order phase transition, and the $AB$-adsorbed
phase is similarly separated from the mixed ballistic and $AB$-adsorbed phase
by a first order phase boundary.  The critical point $y_0$ is the solution of
$\lambda(y)-\log\mu_3 = (f{-}g)(\kappa(b)-\log\mu_3)$.}
\label{f9}
\end{figure}

\section{Discussion}
\label{sec:discussion}

We have investigated a self-avoiding walk model of star copolymers where the arms of the star can be of two chemically different types, $A$ and $B$.  The stars interact with a surface at which they can adsorb and $A$ and $B$ vertices have different interactions with the surface.  The star can be pulled off the surface by a force applied either at the central vertex of the star or at a vertex of degree 1.  We have established the free energy dependence as a function of the interaction strengths with the surface and the magnitude of the applied force, and have thus determined the general forms of the phase diagrams. 

In a typical AFM experiment the values of $a$ and $b$ are fixed while $y$ is increased (so that the desorbing force increases).  We illustrate the results of this in figure \ref{f9} for the case where an $A$-vertex of degree 1 is fixed in the surface and the force is applied at another $A$-vertex of degree 1.  In the left hand figure, $b$ is fixed at some value less than $b_c$ so that $B$-arms do not adsorb.  If $a$ is fixed at a value less than $a_c$ the system is in a free phase when $y<1$.  At $y=1$ there is a phase boundary between the free phase and a ballistic phase with free energy  $(2\lambda(y) + (f{-}2) \log \mu_3)/f$.
If $a > a_c$ (so that $A$-arms tend to adsorb), at small forces the $A$-arms will be adsorbed and the $B$-arms will be free.  As $y$ increases the system crosses a phase boundary at $\lambda(y)=\kappa(a)$ into a mixed phase where one $A$-arm is ballistic, the other $A$-arms are adsorbed and the $B$-arms are free.  As $y$ increases further the system reaches a phase boundary at $\lambda(y) - \log \mu_3 = (g{-}1) (\kappa(a) - \log \mu_3)$ and crosses into a ballistic phase.  
Using arguments similar to those used for Figures \ref{f7}(b) and \ref{f8}, making use of Theorem \ref{theo:convex1},  it can be shown that, in the $(\log a, \log y)$-plane, the curved phase boundary of the ballistic phase is concave down towards the ballistic phase (\text{ie} towards the $\log y$ axis), and the curved phase boundary of the $A$-adsorbed phase is concave down towards that phase.

If $b$ is large and fixed, there's a phase where $B$-arms are adsorbed when $a<a_c$ and $y<1$.  As $y$ increases, with $a<a_c$, the system crosses a phase boundary at $y=1$ to a phase where $B$-arms are adsorbed, one $A$-arm is ballistic and the other $A$-arms are free.  Further increasing $y$, the system crosses a phase boundary determined by
$\lambda(y) - \log \mu_3 = (f{-}g)(\kappa(b) - \log \mu_3)$
to a fully ballistic phase.
If $a>a_c$ the system is in an $AB$-adsorbed phase for small $y$ and there is a phase boundary at $ \kappa(a)=\lambda(y)$ to a phase where everything is adsorbed except for one $A$-arm.  Increasing $y$ further gives a transition to the fully ballistic phase.  
In the $(\log a, \log y)$-plane, the curved phase boundary of the ballistic phase is concave down towards the ballistic phase, and the curved phase boundary of the $AB$-adsorbed phase is concave down towards that phase.

We have concentrated on the three dimensional case but our arguments generalize to any dimension $ d \ge 3$.  The situation in $d=2$ is different because one arm of the star can ``shade'' the surface from other arms \cite{Bradly2019a}.  We have not considered the $d=2$ case here.  It would be interesting to extend our treatment to other homeomorphism types such as combs where the branches can be chemically different.  Although we have been primarily interested in copolymeric stars, the system that we have considered also serves as a simple model of mixed micelles \cite{Bhattacharjee}.

\section*{Acknowledgement}
EJJvR acknowledges financial support from NSERC (Canada) 
in the form of a Discovery Grant RGPIN-2019-06303.




\end{document}